\documentclass[a4paper,11pt]{article}
\pdfoutput=1 

\usepackage{jheppub} 
\usepackage{physics}
\usepackage[T1]{fontenc} 
\usepackage{appendix}
\usepackage{physics}
\usepackage{amsmath, amssymb, graphicx}
\usepackage{amsfonts}
\usepackage{stmaryrd}
\usepackage{mathtools}
\usepackage{appendix}
\usepackage{tikz}
\usetikzlibrary{positioning, quotes, fit}
\usepackage{subcaption}

\newcommand{\ben}{\begin{eqnarray}\displaystyle}
\newcommand{\een}{\end{eqnarray}}

\newcommand{\del}{\partial}

\newcommand{\diag}{\hbox{diag}}

\newcommand{\nn}{\nonumber \\}
\newcommand{\mc}{\mathcal}

\DeclareMathOperator{\sdet}{sdet}

\title{Integrability and Conformal Blocks for Surface Defects in $\mathcal{N}=4$ SYM}

\author[a]{Adolfo Holguin}
\author[b]{and Hiroki Kawai}

\affiliation[a]{School of Mathematics, Trinity College Dublin,  17 Westland Row, Dublin 2, Ireland}
\affiliation[b]{Department of Physics, University of California, Santa Barbara, CA 93106, USA}

\emailAdd{holguina@tcd.ie, hkawai@ucsb.edu}
\abstract{We study various aspects of half-BPS surface defect operators in $\mathcal{N}=4$ SYM. For defects on generic points on the moduli space we use superconformal symmetry to fix the form of one-point and two-point functions of half-BPS operators and solve the superconformal Ward identities in terms of superconformal blocks, emphasizing the role of the broken rotational symmetry transverse to the defect in the superconformal block expansion. 
We verify this expansion by the leading-order perturbative calculation for the two-point functions. 
We also investigate the integrability of the defect CFT in the planar limit and argue that the integrability is broken at generic points of the defect moduli. 
The integrability is expected to be restored in the singular point of this moduli space where another "rigid" branch appears, and we provide evidence for this by showing that the defect one-point functions in this case can be mapped to a class of known integrable quenches.  
}

\begin{document} 
\maketitle
\flushbottom

\section{Introduction}

Integrability and holography are powerful tools for understanding the dynamics of strongly coupled large $N$ gauge theories, leading to the solution of the spectral problem of planar maximally supersymmetric Yang-Mills theory \cite{Beisert:2010jr}. At weak coupling, the mixing problem of single trace operators is equivalent to computing the spectrum of an integrable spin chain~\cite{Minahan:2002ve}, mirroring the integrability of the classical sigma model on $AdS_5\times S^5$~\cite{Bena:2003wd}. 
Similar ideas have been adopted for studying the planar limit of the ABJM theory~\cite{Minahan:2008hf, Bak:2008cp}, which is dual to eleven-dimensional supergravity theory $AdS_4\times \mathbb{CP}^3$~\cite{Aharony:2008ug}.  
A more ambitious goal has been to use tools from integrability to compute CFT data beyond the spectrum, such as OPE data. This idea has had fruitful implementations in the context of a certain type of three-point correlators involving determinant operators \cite{Jiang:2019xdz,Yang:2021hrl}, and in defect CFTs (dCFTs) arising from extended operators \cite{deLeeuw:2015hxa, Giombi:2018hsx, Kristjansen:2021abc, Kristjansen:2023ysz} (also refer to~\cite{Kristjansen:2024dnm} for a review). 
In such set-ups the simplicity of the OPE data is closely associated with the integrability of the spin chain (or string) with open boundaries.  These boundaries are often associated with D-branes wrapping maximal cycles at strong coupling, and the OPE data encodes absorption/emission amplitudes of closed string states. 
The integrability of said boundary conditions was first exploited in the context of open string operators \cite{Berenstein:2005vf}, but its role as an integrable boundary state was noticed in the computation of one-point functions in defect CFT. An important tool for this has been the notion of an integrable boundary, or quench, for spin chains \cite{Piroli:2017sei}. Although the connection between integrable boundaries and integrable boundary states is not fully fledged for spin systems, a compelling picture exists connecting integrable matrix product states (MPS) to solutions of the boundary Yang-Baxter equation leading to new interesting integrable quenches. 
This technique has been useful for computing one-point functions of single trace operators with nontrivial boundary conditions both in $\mathcal{N}=4$ SYM and in ABJM, such as various conformal domain wall setups~\cite{deLeeuw:2015hxa, Buhl-Mortensen:2015gfd, deLeeuw:2016ofj, DeLeeuw:2018cal, Kristjansen:2020mhn}, the dCFT associated to a supersymmetric Wilson/'t Hooft loop~\cite{Kristjansen:2023ysz, Jiang:2023cdm}, and cross-cap states~\cite{Caetano:2021dbh}, and also for computing correlation functions involving maximal giant gravitons~\cite{Jiang:2019xdz, Yang:2021hrl} or in the Coulomb branch~\cite{Ivanovskiy:2024vel}. The hope is that this data can be used to set-up a bootstrap program for higher point correlators \cite{deLeeuw:2017dkd, Baerman:2024tql}.

An interesting question is to classify all possible integrable boundaries that appear in $\mathcal{N}=4$ SYM. Most of the work in this direction stems from strong coupling considerations \cite{Mann:2006rh, Dekel:2011ja, Linardopoulos:2021rfq, Linardopoulos:2025ypq}. At weak coupling, determining whether a boundary condition is integrable involves constructing a solution to the corresponding boundary Yang-Baxter equation \cite{DeWolfe:2004zt, Chen:2004mu, Berenstein:2005vf, Hofman:2007xp, Correa:2008av, Correa:2011nz} which in principle requires information about non-planar contributions to the dilatation operator. A more tractable approach is to instead check the selection rules satisfied by the corresponding boundary state and to use said state to construct a solution to the boundary Yang-Baxter equation. 
The precise integrability conditions of the boundary or quench state proposed in~\cite{Piroli:2017sei} are obtained by channel rotation of the boundary Yang-Baxter equation. 
In \cite{deLeeuw:2024qki} a weaker condition was used to identify integrable matrix product states associated to defects in $\mathcal{N}=4$ SYM, reproducing many of the integrable quenches for the $SO(m)$ Heisenberg chain. This was done in an attempt to identify the integrable boundaries enumerated in \cite{Dekel:2011ja}. While the integrable and BPS codimension one and three defects can be identified with the D3-D5 Nahm pole and with Wilson/'t Hooft loops operators, the codimension two cases remain to be studied in detail. The $AdS_3\times S^5$ was argued to correspond to a fuzzy $S^5$ solution to the equations of motion \cite{deLeeuw:2024qki}, which leaves the case of $AdS_3\times S^1$. Expectedly the dual of said integrable D-brane configuration must belong to the class of surface defects introduced in \cite{Gukov:2006jk, Gukov:2008sn}, but it is a priori not clear which ones exactly. 
Part of the goal of this paper is to address this issue. 

The singular solutions describing a half-BPS defect on a two-dimensional surface in $\mathcal N=4$ SYM were first found by Gukov and Witten~\cite{Gukov:2006jk} in order to introduce ramification to the gauge theory description of the geometric Langlands program. 
It is realized by probing D3 branes wrapping $AdS_3\times S^1$ as long as the number of the probe branes is enough smaller than the gauge symmetry rank $N$~\cite{Gukov:2006jk, Drukker:2008wr}. 
The supergravity description dual to the case where the backreaction is non-negligible was found shortly after that~\cite{Gomis:2007fi}. 
The one-point functions of half-BPS local operators on the gauge theory side were computed in~\cite{Drukker:2008wr}, where it was confirmed that the leading-order term in weak-coupling perturbation matches with the highest-order term in the gravity result, i.e. in the strong-coupling regime.
The finiteness of the expansion was further confirmed recently by~\cite{Choi:2024ktc}. 
The Gukov-Witten defects are somewhat peculiar in comparison to the other half-BPS supersymmetric defects. 
One important difference is the explicit dependence of the field profiles on the angular direction transverse to the defect due to the twisting of the spacetime and $R$-symmetries. 
This somewhat unusual feature results in correlation functions that have explicit dependence on the transverse angular direction for scalar operators of generic $R$-symmetry polarizations. 
This also has important consequences for the superconformal block decomposition of two-point correlation functions \cite{Liendo:2016ymz}. We elaborate on this point by solving the superconformal Ward identities for one-point and two-point functions involving chiral primaries. This is a first step towards setting a bootstrap program for surface defects in $\mathcal{N}=4$ SYM. To our knowledge this has not been elaborated on in the literature. Another feature is the appearance of continuous moduli; while most of the integrable defects known depend on discrete data such as the rank of the MPS, the GW defect admits a continuous moduli space parametrized by a set of central charges $(\mathfrak{Z}, \alpha, \eta)$. 
Most of the known integrable boundary states do not depend on continuous parameters\footnote{A notable exception are the single site states relevant for Coulomb branch operators \cite{Ivanovskiy:2024vel}.} and indeed we find evidence for the breaking of integrability for generic points in the moduli space. We identify the integrable defect with a special point of the moduli space $\alpha = \mathfrak{Z}=0$ at which the fields develop logarithmic behaviour signaling spontaneous breaking of scaling symmetry \cite{Gukov:2008sn}. 
The defects at this singular point are rigid  and the corresponding boundary state describes a fuzzy circle much like the other defects described by fuzzy spheres. 
Note that by rigid we simply mean that they do not depend on continuous parameters, rather than the rigidity condition required by \cite{Gukov:2008sn}. Their condition is more strict in the sense that their conjugacy classes cannot be continuously deformed to nearby conjugacy classes. 
For $G = SU(N)$, the rigid surface defect we consider here obtained by setting $\alpha=\beta=\gamma =0$ has the conjugacy class as the limit of that of the generic case, i.e. its hyperkähler moduli of the 2d Higgs branch in the corresponding 4d-2d description is nontrivial and hence the defect is non-rigid in their definition\footnote{\cite{Gukov:2008sn} found the operators with rigid conjugacy classes for the $SO(N)$ and $Sp(N)$ gauge theories. There is a difference between $SO(2n)$ groups and $SO(2n+1)$ groups since the former is self-dual under S-duality while the latter is dual to $Sp(n)$. The difference comes from the existence of "discrete torsion" in the holographic description~\cite{Witten:1998xy}.  }.  
They are also singled out by the appearance of an additional dilatation mode associated to the spontaneously broken scaling symmetry. 

Our paper is organized as follows. In section \ref{sec:bps surface defects}, we review the disorder defect description of the Gukov-Witten defects, and comment on their holographic description. In section \ref{sec:Ward identities}, we derive the superconformal Ward identities relevant for the two-point functions of bulk chiral primaries. We use these in section \ref{sec: BPS correlators} to determine the form of the superconformal block expansion of bulk two-point functions, and we also constraint the form of one-point function and bulk-to-defect two-point functions. In section \ref{one-point functions}, we study one-point functions of non-protected operators at leading order in perturbation theory. We find that although scalar operators can be associated to an integrable boundary state, this integrability does not persist to sectors with spin due to the broken rotational symmetry transverse to the defect. This symmetry is restored in the rigid case and the resulting integrable states are analogous to those relevant for the D3-D5 interface and the maximal giant graviton. In section \ref{sec:propagators}, we review the quantization of the theory around a generic (non-rigid) defect by mapping it to a theory on $AdS_3\times S^1$.  In section \ref{sec: two pts}, we compute the leading-order contribution to the two point function of chiral primaries show that it organizes according to our superconformal block expansion by identifying the exchanged modes. Finally, we conclude and comment on future directions.

\section{BPS Surface Defects}\label{sec:bps surface defects}
Before discussing the details of the Gukov-Witten defects, let us comment some generalities about defect conformal field theories. 
A defect CFT in $D+1$ dimensions is characterized by coupling a bulk CFT to a lower dimensional conformal theory living on a $p+1$-dimensional defect in a way that preserves the conformal invariance of the defect. 
In Euclidean signature, one usually places the defect along a hyperplane $\mathbb{R}^{p+1}\subset \mathbb{R}^{D+1}$ and the resulting space is conformally equivalent to $AdS_{p+2}\times S^{D-p-1}$:
\begin{equation}
    ds^2= \delta_{\mu 
\nu}dx^\mu dx^{\nu} +dr^2 +r^2 d\Omega^2= r^2 \left(\frac{\delta_{\mu 
\nu}dx^\mu dx^{\nu}+dr^2}{r^2}+ d\Omega^2\right)
\end{equation}
In the new conformal frame, the defect is placed at the boundary of $AdS$, while the original CFT lives in the bulk of the spacetime. 
The same coordinate transformation can be done in a Poincar\'e patch of the Lorentzian cylinder $\mathbb{R}\times S^{D}$, and the coordinates can be extended to global $AdS$ coordinates. 
This formulation is useful since it makes the residual conformal symmetry manifest as the spacetime isometry, and it gives a natural language for describing correlation functions in terms of Witten diagrams. 
Because the full conformal symmetry is partially broken, the operators of the bulk theory can have non-vanishing one-point functions. 
These nonzero one-point functions translate to nonzero asymptotic values of the field operators near the boundary of $AdS$. 
This implies that certain bulk operators can be associated with boundary operators by expanding the operator near the boundary:
\begin{equation}
    \mathcal{O}_{\text{bulk}}( r, \vec{x}, \Omega)=\frac{1}{r^{\Delta}}\sum_{\hat{\mathcal O}}\mathsf{b}_{\mathcal O \hat{\mathcal O}} r^{\hat \Delta} \hat{D}(r,\Omega, \partial_\mu)\hat{\mathcal{O}}_{\text{bdy}}(\vec{x},\Omega)+ \dots,
\end{equation}
where  $\hat{D}$ are differential operations determined by the boundary conformal symmetry. This is the usual defect operator product expansion. 
For general codimensions, one can also expand the dependence of the operator on the normal coordinates to the defect by performing a mode expansion on $S^{D-p-1}$:
\begin{equation}
   \hat{\mathcal{O}}_{\text{bdy}}(\vec{x},\Omega)= \sum_{J} \frac{x_\perp^{i_1}\dots x_\perp^{i_J}}{r^J}\,\hat{\mathcal{O}}_{\text{bdy}}^{i_1\dots i_J}(\vec{x}). 
\end{equation}
The resulting operators in the expansion are primary operators of the boundary (defect) theory.
The leading term in the expansion $\mathsf{b}_{\mathcal O \hat {\mathbb{I}}}= \mathsf{a}_{\mathcal O}$ encodes the one-point functions of bulk operators, or equivalently the coupling to the identity operator of the defect theory. 
These new couplings are additional data necessary to specify the defect operators, and their values are generically not determined by conformal symmetry alone. 
However, their values are not completely independent of the conformal data of the bulk and defect due to constraints arising from the crossing equations.  These constraints are a consequence of the fact that the two-point function of bulk primaries has two different expansions depending on whether the operators are first fused with the bulk OPE or first decomposed with the boundary-defect operator expansion. 
In either case, the kinematic dependence of the correlator is not fully fixed by conformal invariance, but the result can be expanded into bulk and defect conformal partial waves which give an infinite set of equations relating bulk OPE coefficients to bulk-to-defect couplings. 

The class of surface defects we will focus on are those introduced in \cite{Gukov:2006jk, Gukov:2008sn}, although there are other maximally supersymmetric surface operators for the $\mathcal{N}=4$ SYM theory. 
For a comprehensive review of maximally supersymmetric defects of general codimension, see \cite{Wang:2020seq}. 
This class of surface defects differs from the supersymmetric defects of codimension one and three in that they are generically parametrized by a continuous moduli. 
The coordinates of this moduli space are a set of central charges $(\mathfrak{Z}_I, \alpha_I,\eta_I)$. 
The parameters $\alpha_I$ and $\eta_I$ are circled valued, and they encode information about the monodromy of line operators in the presence of the surface operator. 
These cannot be measured by the insertion of local operators, so they will not play a prominent role in our discussion. 
The remaining complex coordinate $\mathfrak{Z}_I$ parametrizes the expectation value of certain scalar operators. 
One well-studied observable is the one-point function of chiral primary operators in the presence of a surface defect \cite{Drukker:2008wr,Choi:2024ktc}. 
Such correlators are protected by supersymmetry and can be computed using techniques from supersymmetric localization; they have also been studied at strong coupling using holographic methods. Some results on correlators of the stress tensor $T^{\mu \nu}$ and the displacement operators $\mathbf{D}_{z, \bar{z}}$ in more general $\mathcal{N}=2$ theories have been obtained \cite{Bianchi:2019sxz} from their associated chiral algebras. 
Two-point functions of chiral primaries with surface defects have not been studied in detail, unlike the codimension-one case where a bootstrap program has been proposed \cite{Liendo:2016ymz}, mainly due to the fact that the residual conformal symmetry of the Gukov-Witten is somewhat unconventional in that the transverse rotational symmetry is broken, which allows for a richer structure for correlation functions. 
More precisely, the residual symmetry of the system is $(PSU(1,1|2)\times PSU(1,1|2))\ltimes SO(2)_t$, where the $SO(2)_t$ symmetry is a diagonal subgroup of the product $SO(2)\times SO(2)_R\subset SO(2,4)\times SO(6)$. In non-supersymmetric set-ups, the residual $SO(2)$ symmetry plays an important role in fixing the dependence of correlation functions of bulk operators on the angle transverse to the defect~\cite{Billo:2016cpy}. For instance, the one-point functions of operators carrying transverse spin vanish in those cases, which is not the case for this Gukov-Witten dCFT. This also makes the defect conformal block expansion of bulk two-point functions  more intricate.

\subsection{Gukov-Witten Defect: Generic Case\label{sec:GW-defect-generic}}
The Gukov-Witten defect~\cite{Gukov:2006jk, Gukov:2008sn} is an extended operator of the $\mathcal{N}=4$ SYM theory defined on a codimension two surface. We will work with the theory in flat space $\mathbb{R}^4$ placing the defect along a two dimensional plane $\mathbb{R}^2 \subset \mathbb{R}^4$. Often we will use the complex coordinate $z$ to denote the holomorphic coordinate normal to the defect. 
The defect can be described by a singular boundary condition on the fields of the theory, we will concentrate on the case in which the defect preserves one-half of the supersymmetries which corresponds to the breaking of the full $PSU(2,2|4)$ symmetry to $(PSU(1,1|2)\times PSU(1,1|2))\ltimes SO(2)_t$\footnote{Another way of describing this surface defect is by coupling the two-dimensional $\mc N = (4,4)$ superconformal sigma model or quiver gauge theory to the four-dimensional theory. The target space of this sigma model is the moduli space of the solutions of Hitchin's equations. 
This 4d-2d description is studied in detail in~\cite{Gukov:2006jk, Gadde:2013dda, Gaiotto:2013sma}. }. 
The details of the embedding of the residual symmetry to the bulk symmetry are discussed in the next section. 
The bosonic part of this residual symmetry is 
$SO(2,2)\times SO(4)_R\times SO(2)_t$.
The two copies of $PSU(1,1|2)$ respectively contain the holomorphic and antiholomorphic parts of the two-dimensional conformal symmetry $SO(1,3)$, as well as they contain each of the left and right $SU(2)$ components of $SO(4)_R$. 
The $SO(2)_t$ symmetry is the diagonal part of the direct product of the spacetime and R-symmetry rotations in the transverse directions. 
Vanishing of the flux of the supercurrent through the defect imposes the BPS equations:
\begin{equation}
\begin{aligned}
    F_{z\bar{z}} + [Z,\bar{Z}]&=0  \\
    D_{\bar{z}} Z&=0,
\end{aligned}
\end{equation}
where $Z=\phi_1+i \phi_2$\footnote{We take the convention where all the adjoint fields take Hermitian values.}.
They are exactly Hitchin's equations obtained by dimensional reduction of the parallel directions to the defect. 
Solutions consistent with classical scale invariance are of the form
\begin{equation}
\begin{aligned}\label{eq:generic-vev}
    Z&= \frac{\phi_1+ i \phi_2}{\sqrt{2}}= \frac{\beta + i \gamma}{z}\\
    A_\psi&= \alpha d\psi,
\end{aligned}
\end{equation}
where $z$ is the coordinate normal to the defect and $z=r e^{i \psi}$. 
We choose a specific gauge where the part of the gauge field $A$ proportional to $dr$ vanishes. 
The parameters $\alpha, \beta ,\gamma$ are commuting matrices in the Lie algebra $\mathfrak{su}(N)$. 
Specifically, they take values in either the Cartan subalgebra $\mathfrak t$ of $\mathfrak{su}(N)$ or the maximal torus $\mathbb T$: 
\begin{align}
    (\alpha, \beta, \gamma) \in (\mathbb T, \, \mathfrak t, \, \mathfrak t)/W_{\mathbb L}
\end{align}
$\mathbb L = S(\prod_{l = 1}^M U(N_l))\; (\sum_{l = 1}^M N_l = N)$ is the Levi subgroup of $SU(N)$, and the gauge symmetry breaking pattern is specified by the values of the diagonal elements of $\alpha,\beta, \gamma$ to be $SU(N) \rightarrow  \mathbb L$. 
We take the quotient by the Weyl group $W_{\mathbb L}$ to account for the equivalence under the exchanges of the Cartan elements having the same values in $\alpha, \beta, \gamma$. 
This is a consequence of the BPS equations, and the description is semiclassical.
In other words, the insertion of the surface operator leads to a new saddle of the path integral; these are classical solutions to the equations of motion with the correct charges. 
The quantum description of the surface operator is as a boundary condition for the field operators at the location of the defect in the path integral. 
There is also an additional parameter $\eta$, taking a value in the maximal torus of the Langlands dual of the gauge group, associated with theta terms on the defect but they will not be important for our discussion.  

\subsubsection{Holographic dual of the generic case}
The Gukov-Witten defect is holographically dual to a collection of $M$ D3 branes wrapping an $AdS^3\times S^1$ inside of $AdS_5\times S^5$. 
There are two ways of describing these D3 branes, either as probes or as a solution to type IIB supergravity. 
To describe the probe, which is valid as long as $M \ll N$, one chooses a slicing of $AdS_5\times S^5$:
\begin{eqnarray}
    ds^2= \frac{\cosh^2 u}{r^2}\left(- dt^2+dx_1^2+dr^2\right)+ du^2+ \sinh^2 u d\psi^2 + d\Omega_5^2.
\end{eqnarray}
The branes sit at a constant non-zero value of $u_0$ and at an equator of the $S^5$ with coordinate $\phi=\phi_0-\psi$. The induced metric on the brane is that of an $AdS_3\times S^1$ of size $\cosh u_0$, and the moduli of the defect are identified with the position of the probes as
\begin{equation}
    \beta+ i \gamma= \frac{\sqrt{\lambda}}{4\pi}\sinh u_0 \,e^{i \phi_0}.
\end{equation}
When the number of branes is comparable to $N$, the more appropriate description is in terms of a backreacted background. The dual supergravity solutions were found in~\cite{Gomis:2007fi} by analytic continuation of the LLM solutions~\cite{Lin:2004nb}:
\begin{equation}
\begin{aligned}
    ds^2&= \mathcal{H}^{-2}\left[\left(\zeta+\frac{1}{2}\right) ds_{AdS_3}^2+ \left(\zeta-\frac{1}{2}\right)d\Omega_3^2 +(d\psi + V)^2\right]+ \mathcal{H}^2(dy^2 + d\xi_i^2)\\
    \mathcal{H}&= \frac{\sqrt{\zeta^2-\frac{1}{4}}}{y}.
\end{aligned}
\end{equation}
The metric is fully fixed in terms of a single function $\zeta(\xi_i, y)$ whose domain is $\mathbb R^2\times \mathbb R_+ $, satisfying an $SO(4)$ symmetric Poisson's equation in six-dimensions. 
The boundary conditions considered in \cite{Gomis:2007fi} are specified in terms of a set of $M$ point charges in this three-dimensional base space. 
The coordinate $\psi$ smoothly degenerates at these points, and the local geometry is that of a Taub-NUT space around the sources. 
Note that this is qualitatively different from the boundary conditions for LLM geometries which are specified by a boundary condition at $y=0$ where different $S^3$ cycles degenerate smoothly. 
The resulting space is smooth and is a result of fibering $AdS_3\times S^3 \times S^1$ on a three-dimensional base space $\mathbb R^2 \times \mathbb R_+$. 
The isometry of the fiber is exactly the bosonic part of the residual symmetry $SO(2,2)\times SO(4)\times SO(2)$. 
The gauge symmetry breaking of $\mathcal N=4$ SYM theory is characterized by the point charge sources in $\mathbb R^2 \times \mathbb R_+$, where the $S^1$ direction becomes degenerate. 
More specifically, the dimension of each of the diagonal blocks of the Levi subgroup corresponds to the position of the particle in $\mathbb R_+$. 
The Higgs scalar VEVs are exactly the positions in $\mathbb R^2$. 
$\alpha$ and $\eta$ correspond to the holonomies of two-form NS-NS and R-R gauge fields around the disks starting from each of the particles and ending on the asymptotic boundary.

\subsection{Residual Symmetry}
Now we review aspects of the residual symmetry $PSU(1,1|2)^2\ltimes SO(2)_t$. 
The four fermionic generators for each of the $PSU(1,1|2)$ groups are explicitly written down for example in \cite{Wang:2020seq}. 
They can be found by solving the Killing spinor equations $\Gamma^{1245}\epsilon_\pm = \epsilon_\pm$ regarding the supersymmetry equations for this half-BPS configuration. 
We follow the notations in~\cite{Wang:2020seq} for the bulk $PSU(2,2|4)$ generators to express the residual symmetry generators, except the R-symmetry charges are indexed with the $SO(6)$ vector indices $I = 1,2,...,6$, instead of the 10d notations\footnote{Here we define $Q (\Tilde Q)$ and $S (\Tilde S)$ as the (anti)chiral generators in terms of the 4d chiral operator $\gamma^5$, and the $Q$ operators and $S$ operators are distinguished in terms of the 10d chiral operator $\Gamma^{11}$. }. 
The fermionic generators of the two copies of $\mathfrak{psu}(1,1|2)$ in terms of the full bulk algebra $\mathfrak{psu}(2,2|4)$ are 
\begin{align}\label{supercharges1}
    \hat Q_{1 a} &= \frac{1}{\sqrt{2}}(Q_{1a \dot a} - i {(\sigma^1)^{\dot b}}_{\dot a}  Q_{1a \dot b}),\quad  
    \hat{\bar Q}_{1 a} = \frac{1}{\sqrt{2}}(\bar Q_{1a \dot a} - i {(\sigma^1)^{\dot b}}_{\dot a}  \bar Q_{1a \dot b}), \nn  
    \hat S_{1 a} &= \frac{1}{\sqrt{2}}(S_{2a \dot a} - i {(\sigma^1)^{\dot b}}_{\dot a}  S_{2a \dot b}),\quad  
    \hat{\bar S}_{1 a} = \frac{1}{\sqrt{2}}(\bar S_{2a \dot a} - i {(\sigma^1)^{\dot b}}_{\dot a}  \bar S_{2a \dot b})
\end{align}
and 
\begin{align}\label{supercharges2}
    \hat Q_{2 a} &= \frac{1}{\sqrt{2}}(Q_{2a \dot a} + i {(\sigma^1)^{\dot b}}_{\dot a}  Q_{2a \dot b}),\quad  
    \hat{\bar Q}_{2 a} = \frac{1}{\sqrt{2}}(\bar Q_{2a \dot a} + i {(\sigma^1)^{\dot b}}_{\dot a}  \bar Q_{2a \dot b}), \nn 
    \hat S_{2 a} &= \frac{1}{\sqrt{2}}(S_{1a \dot a} + i {(\sigma^1)^{\dot b}}_{\dot a}  S_{1a \dot b}),\quad  
    \hat{\bar S}_{2 a} = \frac{1}{\sqrt{2}}(\bar S_{1a \dot a} + i {(\sigma^1)^{\dot b}}_{\dot a}  \bar S_{1a \dot b})
\end{align}
We choose the basis with $a = \dot a$ on the right-hand side. 
We further package these odd generators together with the bosonic generators on a new basis as 
\begin{align}
    \left(\begin{array}{c|c}
        {(\mathfrak{L}_i)^{\alpha}}_{\dot \alpha} & {(\mathfrak{Q}_i)^\alpha}_a  \\
        \hline
        {(\mathfrak{S}_i)^{\dot a}}_{\dot\alpha} &  {(\mathfrak{R}_i)^{\dot a}}_{a}
    \end{array}
    \right) 
    , \quad 
    {(\mathfrak{Q}_i)^\alpha}_a = \begin{pmatrix}
        \hat Q_{i1} & \hat Q_{i2} \\ \hat {\bar S}_{i1} & \hat{\bar S}_{i2}
    \end{pmatrix}, 
    \quad 
    {(\mathfrak{S}_i)^{\dot a}}_{\dot \alpha} = \begin{pmatrix}
        \hat S_{i2} & -\hat {\bar Q}_{i2} \\ \hat S_{i1} & -\hat{\bar Q}_{i1}
    \end{pmatrix}
\end{align}
The bosonic generators of $\mathfrak g_0 = \mathfrak{su}(1,1)\oplus\mathfrak{su}(2) \subset \mathfrak{psu}(1,1|2)$ are, in terms of the full $\mathfrak{psu}(2,2|4)$ algebra generators, 
\begin{align}\label{bosoniccharges}
    {(\mathfrak{L}_i)^{\alpha}}_{\dot \alpha}
    &= \frac{1}{2}\begin{pmatrix}
        \pm D + i M_{34} & - 2P_{1 \dot 1} \\ 
         2K_{2 \dot 2} & - (\pm D + i M_{34})
    \end{pmatrix}, \nn
    {(\mathfrak{R}_i)^{\dot a}}_a
    &= \frac{1}{2}\begin{pmatrix}
    i (\pm R_{62} - R_{14}) &  -(R_{43} \pm R_{56} + i (R_{53} \mp  R_{46})) \\ 
    (R_{43} \pm R_{56} + i (R_{53}\mp R_{46})) &  - i (\pm R_{62}-  R_{14})
    \end{pmatrix}
\end{align}
The signs on top are for the first copy $i = 1$ and the ones on bottom are for the second copy $i = 2$. 
One can verify the $\mathfrak{psu}(1,1|2)$ algebra in this basis as 
\begin{align}
    &\{{(\mathfrak{Q}_i)^\alpha}_a, {(\mathfrak{Q}_i)^{\beta}}_{b}\} = 0, \quad 
    \{{(\mathfrak{S}_i)^{\dot a}}_{\dot \alpha}, {(\mathfrak{S}_i)^{\dot b}}_{\dot \beta}\} = 0 \nn 
    &
    \{{(\mathfrak{Q}_i)^\alpha}_a, {(\mathfrak{S}_i)^{\dot b}}_{\dot \beta}\}
    = \delta^{\dot b}_a {(\mathfrak{L}_i)^{\alpha}}_{\dot \beta} + \delta^{\dot \beta}_\alpha {(\mathfrak{R}_i)^{\dot b}}_{a} + \delta^{\dot b}_a \delta^{\dot \beta}_\alpha \mathfrak{C}
\end{align}
This choice of the basis we made is for being consistent with the notations for $\mathfrak{psu}(2|2)$ in~\cite{Beisert:2006qh}. 
$\mathfrak{C}$ is the generator of the central extension $SO(2)_t$ common to both algebras, which is identified to be $i(M_{12} + R_{12})$. 
Representations of this algebra can be constructed using two copies of representations of $\mathfrak{sl}(2|2)$ by setting their central charges to be equal to each other. More details on the representations of the residual algebra are given in appendix~\ref{sec:residual-reps}.
One important feature of the algebra $\mathfrak{psl}(2|2)$ is that we can turn on two other central extensions besides $\mathfrak{C}$:
\begin{equation}
\begin{aligned}
    &\{{(\mathfrak{Q}_i)^\alpha}_a, {(\mathfrak{Q}_i)^{\beta}}_{b}\} = \epsilon^{\alpha \beta} \epsilon_{a b}\mathfrak{P},\\
    &\{{(\mathfrak{S}_i)^{\dot a}}_{\dot \alpha}, {(\mathfrak{S}_i)^{\dot b}}_{\dot \beta}\} = \epsilon_{\dot \alpha \dot \beta} \epsilon^{\dot a \dot b} \mathfrak{K}. 
\end{aligned}
\end{equation}
The two additional extensions play a key role in determining the asymptotic spectrum of closed string states and of open strings attached to giant gravitons \cite{Beisert:2006qh,Berenstein:2014zxa}. The centrally extended algebra has an additional $\mathfrak{sl}(2, \mathbb{C})$ outer automorphism that can be used to relate the representations of the algebra without central extensions to the centrally-extended case. For the defect algebra this outer automorphism acts as $SU(2)_a$ transformation that leaves the norm of the vector $(\mathfrak{C}, \mathfrak{P}, \mathfrak{K})$ invariant. 
For the surface operator, the defect moduli $(\mathfrak{Z}, \mathfrak{Z}^\dagger)$ plays the role of these central extensions. The BPS representations of the centrally extended algebra satisfy shortening conditions involving $\mathfrak{C}^2+|\mathfrak{P}|^2 $ whose eigenvalues we can associate to $\ell^2+ |\mathfrak{Z}_{ij}|^2$, where $\ell$ is the spin of the operator along the transverse directions to the defect. This leads to composite defect operators with dimensions 
\begin{equation}
    \hat{\Delta}+S= \sum_{i,j}\sqrt{\ell^2+ |\mathfrak{Z}_{ij}|^2}.
\end{equation}
We will see that these are precisely the off-diagonal fluctuations of the $\mathcal{N}=4$ SYM fields.  The physical operators do not carry the central charge $\mathfrak{P}$ due to gauge invariance.

\subsection{Singular Limit $\alpha, \beta, \gamma \rightarrow 0$: Fuzzy Circle Solutions}
The insertion of the surface operator is supposed to introduce singularities for some of the fields, but naively taking the limit $\alpha, \beta, \gamma \rightarrow 0$ leaves no divergent fields. 
This naive assumption is false given that the moduli space has an additional branch corresponding to another family of less divergent solutions for the BPS equations~\cite{Gukov:2006jk, Gukov:2008sn}. 
We can rewrite the BPS equations by introducing a one-form $\phi= Zdz+ \bar{Z} d\bar{z}$ and the field strength $F= F_{z\bar{z}} dz \wedge d\bar{z}$, and the covariant derivative $d_A = d-i A$ as
\begin{equation}
\begin{aligned}
    F -i \phi\wedge \phi&=0  \\
   d_A \phi&=0\\
   d_A\star \phi&=0 .
\end{aligned}
\end{equation}
One can assume the general form of the solutions with rotational invariance in the transverse directions to the defect as 
\begin{equation}
\begin{aligned}   
    \phi&= a_1(r) \frac{dr}{r} + a_2(r) d\psi\\ 
     A&= a_3(r) d\psi\\ 
\end{aligned}
\end{equation}
We again choose a gauge so that the $dr$ part of the gauge field $A$ vanishes. 
Changing variables to $\rho= -\log(r/\Lambda)$, the BPS equations reduce to a set of Nahm equations
\begin{equation}
\begin{aligned}
   \frac{da_a}{d\rho}&= -i\epsilon^{abc}[a_b,a_c].\\
\end{aligned}    
\end{equation}
The field profile is related to the one-point function of the relevant operator in the defect background, so we usually assume that $a_i$ are constants to be consistent with conformal symmetry. 
This assumption is exactly what we made for the generic case we considered in Sec.~\ref{sec:GW-defect-generic}. 
However, solutions with $\rho$ dependence are still consistent with conformal symmetry, provided the scale $\Lambda$ is treated as a zero mode. These solutions are of the form 
\begin{equation}
    a_i= \frac{t_i}{\rho},
\end{equation}
where $t_i$ satisfy an $SU(2)$ algebra. 
Now, field profiles contain an additional logarithmic behavior near $r=0$. This multiplicative factor can be interpreted as an additional dilatation mode of the defect. Naively this field profile breaks conformal invariance due to the appearance of a cut-off scale $\Lambda$, but the value of $\Lambda$ is unphysical since it can be set to any value by using scaling transformations. A more precise statement is that the field profile of the scalar field depends on the zero value of a dilaton mode $\chi_0(r)= 1/\log(r/\Lambda)$
\begin{equation}
    Z^{cl}= \frac{e^{-i\psi}}{2r}(t_1 - it_2)\,\chi_0(r)
\end{equation}
This mode appears as a Goldstone boson for the spontaneously broken scaling symmetry.  
As explained in \cite{Gukov:2008sn}, the correct treatment in this case is to quantize the theory in the presence of this dilaton mode.
In other words, including the fluctuations of this field restores the scaling symmetry.
In practice, this means that we should couple the theory to a dynamical dilaton field which replaces the derivatives with respect to $r$ with a Weyl connection \cite{Monin:2016jmo}. 
The logarithmic divergence alleviates the leading singularity in the $r\rightarrow 0$ limit, so these solutions appear near the defect only when the leading divergent terms disappear. The logarithmic behavior of the correlators should be interpreted as operator mixing due with the dilaton.
Now the surface defect is labeled by a representation of $SU(2)$ and a map $\pi:\;SU(2)\rightarrow U(N)$. We will refer to this class of defects as \textit{rigid} since they do not depend on continuous moduli. 

One important distinction of these solutions to the generic defect is that the $SO(2)_R$ symmetry is no longer broken. 
The scalar field profiles near the defect are given by 
\begin{equation}
\begin{aligned}
     \phi_1^{cl}&= \frac{1}{r \log r/\Lambda} \left[t_1 \cos \psi - t_2 \sin \psi\right]= \chi_0(r)\frac{\mathcal{U}_\psi^\dagger t_1 \mathcal{U}_\psi }{r} =\chi_0(r)\tilde{\phi}_1^{cl}\\
     \phi_2^{cl}&= \frac{1}{r \log r/\Lambda} \left[ t_1 \sin \psi + t_2 \cos \psi\right]= \chi_0(r)\frac{\mathcal{U}_\psi^\dagger t_2 \mathcal{U}_\psi }{r} =\chi_0(r)\tilde{\phi}_2^{cl} \\
A_\psi^{cl}&=\chi_0(r) t_3= \chi_0(r) \tilde{A}_\psi^{cl},\;\;\;\;
\mathcal{U}_\psi= e^{-i t_3\psi }.
\end{aligned}
\end{equation} 
Note that the "bare" scalars $\tilde{\phi}_{1,2}^{cl}$ are covariantly constant with respect to $\partial_\psi- i \tilde{A}^{cl}_{\psi}$, which means that we restore the rotational symmetry in $\psi$ by turning on a background gauge. Additionally, since the scalar VEVs satisfy the algebra of $SO(3)$ along with $A_\psi$, we can think of them as coordinates on a fuzzy circle. 
Strictly speaking, we have a collection of fuzzy circles whose radii depend on the eigenvalue of $t_3$.
Clearly, there is a symmetry $SO(2)$ generated by $t_3$ that rotates $\phi_{1,2}$ leaving their relations fixed. 
For the generic defect this symmetry is an outer automorphism that maps inequivalent defects to each other, but at the origin of the moduli space this outer automorphism becomes a symmetry, untwisting the transverse symmetry.

\subsubsection{Holographic dual of the singular case}
The dual description of these kinds of defect should be qualitatively different from the generic case. Intuitively, one expects that the singular limit corresponds to taking $u_0\rightarrow 0$. 
In that case, the $\psi$ direction disappears, and the brane wraps a maximal $AdS_3$ cycle inside of $AdS_5$. On the $S^5$ coordinates, we are free to wrap the remaining direction along the equator giving an induced metric on the brane of $AdS_3 \times S^1$. 
This $S^1\subset S^5$ should be interpreted as a fuzzy circle. 
Since the solution sits at a special point where the $\psi$ coordinate disappears, the rotational symmetry along the $\psi$ coordinate is restored. Moreover, the rotational symmetry along the equator of the sphere is restored since the brane wraps the whole equator. 
The spacetime circle disappears, so we cannot thread NS-NS or R-R 1-form flux, and the defect lacks any continuous theta angles.  
This solution appeared before in \cite{Dekel:2011ja} and is known to correspond to an integrable boundary for the type IIB superstring on $AdS_5\times S^5$.

\section{Superconformal Ward Identities}\label{sec:Ward identities}
In this section, we find the superconformal Ward identities (SCWIs) for the two-point functions of half-BPS scalar operators in the generic background.
Finding these symmetry constraints is required for fully determining the superconformal blocks composing the two-point function. 
We utilize the harmonic superspace formalism, which is set up for this Gukov-Witten defect by~\cite{Liendo:2016ymz}, for a simple comprehensive expression of the superconformal transformations of the bosonic and fermionic coordinates. 
This superspace formalism approach for finding SCWIs is first considered in~\cite{Dolan:2004mu} for the four-point function of the $\mc N=4$ theory, and also considered for the codimension-one defect case in~\cite{Liendo:2016ymz}. 
The harmonic superspace formalism and the action of the residual $PSU(1,1|2)^2 \ltimes SO(2)_t$ symmetry on it is reviewed in appendix~\ref{sec:superspace}. 
The spacetime coordinates $x^\mu$ and the R-symmetry space polarization $Y^I$ can be packaged as a $(2|2)$ matrix together with the Grassmann odd coordinates as 
\begin{align}
    X^{A\dot A} = \begin{pmatrix}
        x^{\alpha\dot \alpha} & \theta^{\alpha \dot a}\\
        \bar \theta^{a \dot \alpha} & y^{a\dot a}
    \end{pmatrix}, \quad A,\dot A = 1,...,4, \; a,\dot a, \alpha, \dot \alpha = 1,2. 
\end{align}
Let us call the harmonic superspace coordinates of the two operator-insertion points $X_1$ and $X_2$. 
$X$'s can be split to the parallel components and to the transverse components with respect to the surface defect as $X_i = X_{iS} + X_{i\perp}$. 
The number of the residual supertranslation and superspecial conformal transformation generators is 16 in total, so we can fix the 16 components of $X_{1S}$ and $X_{2S}$ to be zero. 

We need to find the quantity containing the cross-ratios between the two insertion points regarding the residual superconformal symmetry. 
Note that the eigenvalues of the quantity $X_{12\perp} \equiv X_{1\perp} X_{2\perp}^{-1} = 
\begin{pmatrix}\chi_{1\perp}\chi_{2\perp}^{-1}
            & 0\\
            0 & \bar \chi_{1\perp}\bar \chi_{2\perp}^{-1}
\end{pmatrix}$ 
are $PSU(1,1|2)^2 \ltimes SO(2)_t$ invariant. 
Each of the superspace positions $X_1$ and $X_2$ has the eight independent fermionic coordinates $\theta^{\alpha \dot a}$ and $\bar \theta^{a\dot\alpha}$. 
Each group element of the residual symmetry has 16 independent fermionic parameters corresponding to the $Q$ and $S$ generators, which allow us to fix the fermionic coordinates of both $X_1$ and $X_2$ to be all zero (notice this is obviously compatible with the choice $\chi_S = \bar \chi_S = 0$ that we already made). 
This means that the two-point function is fully determined by considering the top components in the supermultiples. 
Hence, the matrices $\chi_{1\perp}$, $\chi_{2\perp}$, and their barred counterparts are diagonal with only bosonic elements. 
The eigenvalues of $X_{12\perp}$ are
\begin{align}\label{eq:x12-evals}
    X_{12\perp}  
    \sim 
    \diag\left(\frac{(\chi_{1\perp})_{11}}{(\chi_{2\perp})_{11}}, \frac{(\bar \chi_{1\perp})_{11}}{(\bar \chi_{2\perp})_{11}},  \frac{(\chi_{1\perp})_{22}}{(\chi_{2\perp})_{22}}, \frac{(\bar \chi_{1\perp})_{22}}{(\bar \chi_{2\perp})_{22}}\right)
    \equiv 
    \diag(u, \bar u, v, \bar v).
\end{align}
The set of these eigenvalues has a one-to-one correspondence\footnote{Namely, the cross-ratios can be expressed in terms of these eigenvalues as 
\begin{align}
    \chi = \frac{u \bar u + 1}{\sqrt{u \bar u}}, \quad 
    \chi_R = \frac{v \bar v + 1}{\sqrt{v \bar v}}, \quad 
    \cos\phi 
    = \frac{u + \bar u}{2\sqrt{u \bar u}}, \quad
    \cos\phi_R
    = \frac{v + \bar v}{2\sqrt{v \bar v}}
\end{align}
when the fermionic coordinates are all zero. }
with the four cross-ratios $\chi$, $\chi_R$, $\phi$, and $\phi_R$ which we define in section~\ref{sec:bulk-bulk-2pt-CB}.
$(\chi_{\perp})_{11}$ coincides with the $ z = x^1 + ix^2 = re^{i\psi}$ coordinate of the transverse directions, and $(\bar \chi_{\perp})_{11}$ coincides with $\bar z$ for the top components with zero odd coordinates. 
Because the Gukov-Witten defect preserves only the diagonal $SO(2)_t$ part of the $SO(2) \times SO(2)_R$ rotations in the perpendicular directions of the spacetime and the R-symmetry space, the two-point functions of the operators charged under the broken part of these transverse rotations have the dependence on the two additional angular variables $\eta_1$ and $\eta_2$: 
\begin{align}\label{eq:eta12-def}
    \cos\eta_1 &\equiv \frac{1}{2}\left(\left(\frac{\sdet(\chi_{1\perp})}{\sdet(\bar \chi_{1\perp})}\right)^{1/2} + \left(\frac{\sdet(\bar \chi_{1\perp})}{\sdet( \chi_{1\perp})}\right)^{1/2}\right)
    , \\
    \cos\eta_2 &\equiv \frac{1}{2}\left(\left(\frac{\sdet(\chi_{2\perp})}{\sdet(\bar \chi_{2\perp})} \right)^{1/2} + \left(\frac{\sdet(\bar \chi_{2\perp})}{\sdet( \chi_{2\perp})}\right)^{1/2}\right). 
\end{align}
They are invariant under the mixed $SO(2)_t$ rotations but not under each of the individual $SO(2)$ rotations in the spacetime and the R-symmetry space. 
Actually, there is a combination of the eigenvalues of $X_{12\perp}$ linearly dependent with the combination $\eta = \eta_1 - \eta_2$, so we can say the two-point function depends on only one of the $\eta$ angles besides the eigenvalues.

We write the general expression of the two-point function as
\begin{align}\label{eq:2pt-symmetry}
    &\expval{\mc O_1(X_1) \mc O_2(X_2)}
    = 
    \frac{\mc F_{\mc O_1 \mc O_2}(u, \bar u, v, \bar v, \eta_i)}{\sdet(\chi_{1\perp})^{\Delta_1/2}\sdet(\bar \chi_{1\perp})^{\Delta_1/2}\sdet(\chi_{2\perp})^{\Delta_2/2}\sdet(\bar \chi_{2\perp})^{\Delta_2 /2}}. 
\end{align}
To find the SCWIs, we think of the expansion of this function $\mc F_{\mc O_1 \mc O_2}$ about the fermionic coordinates around $\theta = \bar \theta = 0$. 
We consider the subset of the supersymmetry transformations considered in~\cite{Dolan:2004mu} restricting them to be in the residual symmetry $PSU(1,1|2)^2 \ltimes SO(2)_t$. 
This means that the parameters of the infinitesimal supersymmetry transformations $\epsilon^{\alpha \dot a}$ and $\bar \epsilon^{a \dot\alpha}$ have only diagonal elements.
The infinitesimal transformations of the odd coordinates under this subset of the residual symmetry are
\begin{align}
    \delta \theta^{\alpha\dot a}
    = 
    \begin{pmatrix}
        (v - u) \epsilon^{11} & 0 \\ 0 & (\bar v - \bar u) \epsilon^{22}
    \end{pmatrix}, 
    \quad 
    \delta \bar \theta^{a\dot \alpha}
    = 
    \begin{pmatrix}
        (v - u) \bar \epsilon^{11} & 0 \\ 0 & (\bar v - \bar u) \bar \epsilon^{22}
    \end{pmatrix}, 
\end{align}
and the transformations of the bare eigenvalues $u, \bar u, v, \bar v$ are
\begin{align}
         \delta u = \epsilon^{11}\bar \theta^{11} + \theta^{11} \bar \epsilon^{11}
    , & \quad 
    \delta \bar u = \epsilon^{22}\bar \theta^{22} + \theta^{22} \bar \epsilon^{22}, 
    \nn 
    \delta v = -\bar \epsilon^{11} \theta^{11} - \bar \theta^{11}  \epsilon^{11}
    , & \quad 
    \delta \bar v = -\bar \epsilon^{22} \theta^{22} - \bar \theta^{22}  \epsilon^{22}. 
\end{align}
The linear part of the eigenvalues of $X_{12\perp}$ after the transformation can be seen from these transformation laws as 
\begin{align}\label{eq:evals-SUSY}
    \Tilde u = u - \frac{\theta^{11}\bar \theta^{11}}{u-v}, 
    & \quad 
    \Tilde {\bar u} = \bar u - \frac{\theta^{22}\bar \theta^{22}}{\bar u - \bar v}, \quad 
    \Tilde v = v - \frac{\theta^{11}\bar \theta^{11}}{u - v}, \quad 
    \Tilde {\bar v} = \bar v - \frac{\theta^{22}\bar \theta^{22}}{\bar u - \bar v}. 
\end{align}
The transformation law \eqref{eq:evals-SUSY} suggests that as we expand $\mc F_{\mc O_1 \mc O_2}$ regarded as a function of $(u, \bar u, v, \bar v)$ (plus the $\eta$ variables) around $\theta = \bar \theta = 0$, it encounters a spurious singularity in the R-symmetry space at $v \rightarrow u$ or $\bar v \rightarrow \bar u$. 
This spurious singularity must be suppressed due to the harmonic analyticity of $\mc F_{\mc O_1 \mc O_2}$.
These two limits $v \rightarrow u$ and $\bar v \rightarrow \bar u$ are actually equivalent when we consider the top components, i.e. the case when all of the odd coordinates are zero, since the unbarred and barred variables are complex conjugates of each other. 
The independence of the $(u, v)$ part and the $(\bar u, \bar v)$ part allows us to take care of these singularities separately by considering the SUSY transformation with either $\epsilon^{11} = \bar \epsilon^{11} = 0$ or $\epsilon^{22} = \bar \epsilon^{22} = 0$. 
For these transformations, the change of $\mc F_{\mc O_1 \mc O_2}$ due to the changes of $\eta_1$ and $\eta_2$ is zero since their changes in the linear order are proportional to $\epsilon^{11}\bar \epsilon^{22}$ and $\epsilon^{22}\bar \epsilon^{11}$, and the change in the quadratic order is proportional to $\epsilon^{11}\bar \epsilon^{11}\epsilon^{22}\bar \epsilon^{22}$. 
This makes sense since the $\eta$ dependence is due to the broken part of the twisted symmetry, and the SCWIs are not expected to determine it. 
Let us first consider the case of $\epsilon^{22} = \bar \epsilon^{22} = 0$. 
In this case, only $\theta^{11}$ and $\bar \theta^{11}$ can be the nonzero odd coordinates of $X_{12\perp}$.
Due to the nilpotency of the odd coordinates, the $u, v$ part of the expansion of $\mc F_{\mc O_1 \mc O_2}$ becomes exact with the linear part only: 
\begin{align}
    \mc F_{\mc O_1 \mc O_2}\left(u +\delta u, \bar u, v + \delta v, \bar v, \eta_i\right) 
    = \mc F_{\mc O_1 \mc O_2}(u, \bar u, v, \bar v, \eta_i) - \frac{\theta^{11}\bar \theta^{11}}{u - v} (\del_{u} + \del_{ v})\mc F_{\mc O_1 \mc O_2}(u, \bar u, v, \bar v, \eta_i). 
\end{align}
Hence, the harmonic analyticity of $\mc F_{\mc O_1 \mc O_2}$ at $ v\rightarrow u$ requires the numerator of the linear correction to be zero: 
\begin{align}\label{eq:SCWI-hol}
    (\del_{ u} + \del_{ v})\mc F_{\mc O_1 \mc O_2}(u, \bar u, v, \bar v, \eta_i)\bigg|_{ v \rightarrow u} 
    = 0. 
\end{align}
Similarly, for the other case $\epsilon^{11} = \bar \epsilon^{11} = 0$, the harmonic analyticity requires
\begin{align}\label{eq:SCWI-antihol}
    (\del_{\bar u} + \del_{\bar v})\mc F_{\mc O_1 \mc O_2}(u, \bar u, v, \bar v, \eta_i)\bigg|_{\bar v \rightarrow \bar u}
    = 0. 
\end{align}
These are the superconformal Ward identities for the two-point function.

\section{Half-BPS Correlation functions}\label{sec: BPS correlators}
Here, we determine the general forms of the one-point functions and the two-point functions from the superconformal symmetry. 
We are going to work in the embedding formalism, where the four-dimensional spacetime is embedded in the six-dimensional space by mapping the points in $\mathbb R^{1,3}$ to a light cone $\subset \mathbb R^{2,4}$. 
With this formalism, the conformal symmetry $SO(2,4)$ acts linearly on the coordinates in $\mathbb R^{2,4}$. 
Specifically, the spacetime coordinates $x^\mu$ is mapped to a six-dimensional null vector as 
\begin{align}
    x^\mu \mapsto P^M = (P^\mu, P^+, P^-) = (x^\mu, 1, x^2). 
\end{align}
The insertion of the flat conformal surface defect $\Sigma$ splits this space to 
\begin{align}
    P^M = (P^I, P^A), \qquad I = 1, 2, \quad A = 3, 4, +, -. 
\end{align}
The residual two-dimensional conformal symmetry $SO(2,2)$ now linearly acts as a Lorentz symmetry on $P^A$.
We follow the convention in~\cite{Billo:2016cpy}, where the dot products are denoted with $\cdot$ for the whole six-dimensional vectors, $\circ$ for the transverse directions to the defect, and $\bullet$ for the parallel directions. 
The R-symmetry space coordinates $Y^I$ are split in the same manner as $Y^M = (Y^I, Y^A)$, where the first two "transverse" $I = 1,2$ directions correspond to the real scalars $\phi^1, \phi^2$ with nonzero VEVs. 
The twisting of the transverse spacetime and R-symmetry $SO(2)$ rotations in the residual symmetry defines the dot product between $P^I$ and $Y^I$. 
We denote this also as $P \circ Y$.

\subsection{One-point Functions}
Let us first consider the simplest class of one-point functions. The lowest spin of a half-BPS multiplet is a scalar operator transforming in the symmetric traceless tensor representation of $SU(4)_R$. Like with spinning primaries, we can deal with the $SU(4)_R$ indices by contracting them with an auxiliary null six vector $Y$
\begin{equation}
    \mathcal{O}_{\Delta}(Y,P)= Y_{I_1}\dots Y_{I_{\Delta}}    \mathcal{O}_{\Delta}^{I_1\dots I_{\Delta}}(P).
\end{equation}
One can use the residual conformal symmetry to set the parallel components of the insertion point to zero, and so the one-point function can only depend on the normal coordinates to the defect $z,\bar{z}$. 
The residual $SO(4)_R$ symmetry allows us to set the components of a generic $Y^A$ to be proportional to $(1,0,0,0)$ which leaves a little group $SO(3)$ symmetry. 
Using a conformal transformation, we can further set an additional component of $P^A$ to zero as long as $|z|\neq 0$. In Euclidean signature this would be the Lorentz transformation that brings $P^A$ to a center of mass frame:
\begin{align}
        P^M&\rightarrow \left(z, \bar{z}, 0, 0, 0, |z|\right)\\
        Y^M&\rightarrow(Y_1,Y_2, Y_3,0,0,0).
\end{align}
The residual symmetry left after fixing this frame is $SO(1,2)\times SO(3)$, where both factors are the diagonally embedded Lorentz groups of $SO(2,2)\times SO(4)_R$. The full supersymmetry algebra preserved can be shown to be a diagonal $\mathfrak{psu}(1,1|2)$ of $\mathfrak{psu}(1,1|2)^2$~\cite{Liendo:2016ymz}. 
The one-point function of half-BPS operators is invariant under this subgroup, so only defect multiplets that contain a singlet under this would appear. 
The $SO(4)_R$ symmetry acting on $Y$ can be traded for a transformation acting on the defect which leaves it invariant, so only $SO(4)_R$ singlets are allowed. 
The branching of the half-BPS multiplet in the $SU(4)_R$ representation $[0, \Delta, 0]$ is 
\begin{align}\label{eq:su4-so4so2-half-BPS}
    [0, \Delta, 0] \rightarrow \bigoplus_{ n = 0}^\Delta \bigoplus_{j = -(\Delta/2 - n)}^{\Delta/2 - n} [\tfrac{n}{2}, \tfrac{n}{2}]_{2 j}
\end{align}
where $[\tfrac{n}{2}, \tfrac{n}{2}]_{ m}$ labels the representation of $SO(4)_R \times SO(2)_R$. 
The $SO(4)_R$ invariance requires that the only allowed quantum numbers are $n=0$. 
The spacetime symmetry implies that only defect scalars appear, but there is no constraint for operators carrying transverse spin.
The $SO(2)_R$ spin has no constraint either. 
There is however a constraint coming from the Ward identity for the mixed $SO(2)_t$. 
The only operators with non-zero one-point functions have vanishing central charge under this. 
Putting these constraints all together, the general form of a one-point function of half BPS operators is
\begin{equation}
\begin{aligned}
    \langle O_{\Delta}(Y,P) \rangle&=  \frac{\left(Y\circ Y\right)^{\Delta/2}}{(P\circ P)^{\Delta/2}}\left[\mathsf{a}_{\Delta,0}+\sum_{q=1}^{\Delta/2} \left(\; \frac{(\bar{w} z)^{q}\,\mathsf{a}_{\Delta,q}^{+}+(w\bar{z})^{q}\,\mathsf{a}_{\Delta,q}^{-}}{ \left(Y\circ Y\right)^{q/2}(P\circ P)^{q/2}} \,\right)\right]\\
    &= \frac{\left(Y\circ Y\right)^{\Delta/2}}{(P\circ P)^{\Delta/2}}\times \mathcal{A}_{O}(\eta),
\end{aligned}
\end{equation}
where we introduced $w= Y_1+ i Y_2$ to parametrize the $R$-symmetry polarization. Note that this includes contributions from boundary operators carrying spin in the directions orthogonal to the defect, since the field $\boldsymbol{\phi}$ should not be treated as a scalar with this twisted symmetry $SO(2)_t$. 
We can interpret the one-point function as being fixed not in terms of a single number, but a function $\mathcal{A}_{I}(\eta)$ of the invariant $e^{i 2 \eta}= \bar{w} z/ w \bar{z}$, which specifies the angle between the transverse position vector $P^I$ and the polarization $Y^I$ in this twisted space.

\subsection{Bulk-Defect Two-point Functions}
Bulk operators can have nonzero two-point functions with the defect operators due to their nontrivial OPE decompositions in terms of the defect operators. 
A defect primary is labeled by its quantum numbers under $PSU(1,1|2)^2\ltimes SO(2)_t$. 
Since we will exclusively deal with bulk chiral primaries, the bulk-defect OPE is restricted to have only defect primaries without parallel spin. 
In that case, the quantum numbers under the bosonic symmetries are $\left(\hat{\Delta}, [\frac{n}{2},\frac{m}{2}], \ell\right)$ where $[\frac{n}{2},\frac{m}{2}]$ are $SO(4)_R$ spins. 
In the decomposition of chiral primaries, only $n=m$ representations appear as in~\eqref{eq:su4-so4so2-half-BPS}. 
The symmetry fixing the defect origin and the bulk insertion point is the one fixing the conformal frame, namely $PSU(1,1|2)$. 
So, the only allowed quantum numbers of the defect operators in the OPE correspond to the representations containing a singlet under this stability subgroup. 
We can do this by taking the top component of the multiplet of the $\mathfrak{psu}(1,1|2)^2$ to have $\left([\frac{\hat \Delta}{2},\frac{\hat \Delta}{2}],[\frac{n}{2},\frac{n}{2}]\right)$ quantum numbers under the maximal bosonic subalgebra. 
The shortening conditions for $\mathfrak{su}(1,1|2)$ (see appendix~\ref{sec:residual-reps}) imply the shortening conditions of the residual superconformal symmetry for these defect operators appearing OPE to be $\ell = \pm(\hat \Delta + n)$ or $\ell = \pm(-\hat \Delta + n+2)$. 
The most general defect OPE will involve a sum over the defect supermultiplets.

\begin{equation}
    O_\Delta(P,Y)= \sum_{\hat{O}^{\hat \rho}} \frac{\mathcal{C} (P^A, Y^A, \partial_{\hat{P}}, \partial_{\hat{W}})}{(P\circ P)^{\frac{1}{2}(\Delta-\hat{\Delta})}}\mathcal{F}(z/ \bar{z}, w/\bar w)\, e^{i \ell \psi}\,\hat{O}^{\hat \rho}_{(\hat{\Delta}, n, \ell)}(\hat{P}, \hat{W}),\;\;\; \hat{W}\bullet \hat{W}=0.
\end{equation}
where $\hat{O}^{\hat \rho}$ is the top component of a supermultiplet $\hat \rho$ and $\mathcal{C}$ is a differential operator that takes into account the contribution from superconformal descendants and the hatted vectors are the projections into the defect coordinates.
In our case, $\hat \rho$ are the defect $\mathfrak{psu}(1,1|2)^2\ltimes SO(2)$ representations.
The details of the representations of this Lie superalgebra and their branching to the bosonic components are in appendix~\ref{sec:residual-reps}. 
We also need to introduce an auxiliary complex null vector to take into account the $SO(4)_R$ indices. The dependence in $\psi$ can be understood as follows. 
In the untwisted case, we would have the spacetime $SO(2)$ symmetry in the transverse directions, so the defect operators should be multiplied by additional factors of $(z/\bar{z})^{\ell/2} = e^{i\ell\psi}$ in the OPE, where $\ell$ is the spin in the transverse direction. 
In our case, this symmetry is twisted with the residual R-symmetry, so the correct eigenfunctions on the circle transform as sections of a non-trivial bundle. 
We should think of this as having a constant R-symmetry gauge field on the circle. The most general form this can take is $e^{i \ell \psi}$ times a polynomial in the invariant coordinate $e^{i 2\eta}=\frac{z \bar w}{ \bar{z} w}$.  
Since the only $SO(4)_R$ covariant structure is $Y\bullet \hat{W}$, the bulk-defect two-point function has the general form
\begin{eqnarray}
    \big\langle  O_\Delta(P,Y) \,\hat{O}^{\hat \rho}_{(\hat{\Delta}, n, \ell)}(\hat{P}, \hat{W}) \big\rangle= \frac{(Y\circ Y)^{\frac{1}{2}(\Delta-n)}}{(P\circ P)^{\frac{1}{2}(\Delta-\hat{\Delta})}}  \frac{(Y\bullet \hat{W})^{n}}{(P\bullet \hat{P})^{\hat{\Delta}}}\mathcal{B}_{O, \hat{O}}(\eta). 
\end{eqnarray}
To fix the form of $\mathcal{B}_{O, \hat{O}}$ it is best to expand $O_{\Delta}(P,Y)$ in a basis that is covariant under $SO(2,2)\times SO(2)\times SO(4)_R\times SO(2)_R$ and then perform the defect OPE.
The basis functions are specified by the quantum numbers $(\Delta, j, m, k)$ corresponding to each subgroup of $SO(2,2)\times SO(2)\times SO(4)_R\times SO(2)_R$. 
The most general form of this expansion is
\begin{equation}
\begin{aligned}
   & O_{\Delta}(P,Y) = \sum_{j}\sum_{m=0}^\Delta \sum_{|k|\leq \frac{\Delta-m}{2}} C_{\Delta,m,k} (Y\circ Y)^{\frac{(\Delta-m)}{2}} \left(\frac{w}{\bar w}\right)^{k}\mathcal{D}(Y^A, \partial_{\hat{W}}) e^{ij \psi}O_{(\Delta, m, 2k)}^{(j)}(P,\hat{W}),
    \end{aligned}
\end{equation}
Where $\mathcal{D}$ is an operator accounts for the change of basis between $SO(6)$ and $SO(4)$, and its role is to take into account the contribution of R-symmetry descendants. The same operator must appear in the defect OPE so we do not need its explicit form.  Clearly it is the operators $\tilde{O}_{(\Delta, m, 2k)}^{(j)}(x_\parallel, r,\hat{W})$ with $\ell+j+2k=0$ that contribute to the two point function.  Using this we can deduce that the form of the function is 
\begin{equation}
   \mathcal{B}_{O, \hat{O}}(\eta)= \sum_{|k|\,\leq \frac{\Delta-n}{2}} \mathsf{b}_{O^{(2k-\ell)}_{(\Delta, n,2k)}, \hat{O}_{(\hat{\Delta}, n, \ell)}} \,C_{\Delta, n, k} \,e^{i2k\eta},
\end{equation}
and the coefficients $C_{\Delta,n,k}$ do not depend on defect CFT data and can be determined from bulk branching rules; these can be reabsorbed into the definition of $\mathsf{b}_{O, \hat{O}}$. Just like the one-point function, we can interpret the bulk-defect two-point function as being encoded in a function.

\subsection{Bulk-Bulk Two-point Function\label{sec:bulk-bulk-2pt-CB}}
For the two-point function of bulk operators, superconformal symmetry is not powerful enough to fix the whole correlator. In general there are two ways of expanding the bulk-bulk two-point function, one where we perform the bulk OPE to fuse the two operators and then sum the contributions of each one-point function, and another where each operator is re-written as a sum over boundary operators. Both expressions give different conformal block expansions for the two-point function. We will concentrate on the second expansion, the defect conformal block expansion, for the two-point function of bulk half-BPS operators. 
We start from defining the cross-ratios invariant under the residual symmetry: 
\begin{align}
    &\chi = -\frac{2 P_1 \bullet P_2}{(P_1 \circ P_1)^{1/2}(P_2 \circ P_2)^{1/2}}, \quad  
    \chi_R = -\frac{2 Y_1 \bullet Y_2}{(Y_1 \circ Y_1)^{1/2}(Y_2 \circ Y_2)^{1/2}}\nn
    &\cos\phi = \frac{P_1 \circ P_2}{(P_1 \circ P_1)^{1/2}(P_2 \circ P_2)^{1/2}}, \quad 
    \cos\phi_R = \frac{Y_1 \circ Y_2}{(Y_1 \circ Y_1)^{1/2}(Y_2 \circ Y_2)^{1/2}}.
\end{align}
$\phi$ and $\phi_R$ measures the angles between the position vectors and the polarizations of the two insertion points in the transverse directions, respectively. 
We can fix the frame with the mixed $SO(2)$ rotation so that $\psi_2 = 0$ and $\psi = \psi_1 = \phi$. 
We can start by expanding the $SO(6)_R$ quantum numbers into a mutual $SO(4)_R$ basis. 
By symmetry, the only contributions come from the components of the operator that have the same $SO(4)_R$ quantum numbers, but the $SO(2)_R$ of the bulk theory is no longer conserved. This can be interpreted as having a constant $SO(2)_R$ background gauge field along the $S^1$ normal to the defect just as we mentioned for the bulk-defect two-point function case. As a simple example, let us consider the case of two uncharged chiral primaries under $SO(2)_R$, $O_{\Delta_{1,2}, n_{1,2}, m_{1,2}=0}$. These modes are unaffected by the background gauge field so their two point functions behave as they would if the rotational symmetry normal to the defect was unbroken:
\begin{equation}
\begin{aligned}
 \big\langle O_{(\Delta_{1}, n, 0)}(P_1, \hat{W}_1)O_{(\Delta_{2}, n, 0)}(P_2, \hat{W}_2)\big\rangle&= \frac{(\hat{W}_1\bullet \hat{W}_2)^{n}}{(P_1\circ P_1)^{\Delta_1/2}(P_2\circ P_2)^{\Delta_2 /2}} \,\mathcal{G}(\chi, \psi, \chi_R)   \\
 \mathcal{G}(\chi, \psi, \chi_R)&= \sum_{\{\hat{O}\}} \,\mathsf{b_{1, \hat{O}}} \mathsf{b_{1, \hat{O}}}\, \mathfrak{f}_{\hat{\Delta}}(\chi) \,\mathfrak{f}^R_{[\frac{n}{2},\frac{n}{2}]}(\chi_R) \,\mathfrak{h}_{\ell}(\psi).
\end{aligned}
\end{equation}
where $\mathfrak{f}_{\hat{\Delta}}(\chi),\; \mathfrak{f}^R_{[\frac{n}{2},\frac{n}{2}]}(\chi_R),\; \mathfrak{h}_{\ell}(\psi)$ are the conformal blocks corresponding to each of the representations of the bosonic subgroups, $SO(2,2), SO(4)_R$, and $SO(2)$, respectively. 
Their forms are determined by solving the quadratic Casimir equations for this residual bosonic symmetry as 
\begin{align}
    \mathfrak{f}_{\hat \Delta}(\chi)
    &=  
    \chi^{-\hat \Delta} {}_2 F_1 \left(\frac{\hat \Delta + 1}{2}, \frac{\hat \Delta}{2}; \hat \Delta; \frac{4}{\chi^2}\right) \label{eq:CB-so13}
    \\
     \mathfrak f^R_{[\frac{n}{2},\frac{n}{2}]}(\chi_R)
    &= 
    \chi_R^{n} \, {}_2 F_1 \left(\frac{-n + 1}{2}, -\frac{n}{2}; n; \frac{4}{\chi_R^2}\right)\label{eq:CB-so4R}
    \\
    \mathfrak h_{j}(\psi)
    &= e^{ij\psi }\label{eq:CB-so2}. 
\end{align}
Refer to appendix~\ref{sec:conformal-blocks} for their full derivations. 
For charged operators, the expansion is more complicated. 
This is mainly because there are more invariant quantities than in the case with unbroken transverse rotational symmetry. 
These extra invariants are exactly the $\eta$ angles for each of the insertion points. 
Extending the intuition from the previous correlators, we expect that the most natural way of packaging the defect conformal block expansion is by promoting the couplings to functions $\mathsf{b}\rightarrow\mathcal{B}(\eta)$:
\begin{eqnarray}
    \big\langle O_{\Delta_1} O_{\Delta_2} \big\rangle= \left(\texttt{kinematic}\right) \left[ \mathcal{A}_{1}(\eta_1)\mathcal{A}_{2}(\eta_2)+ \sum_{\hat{\rho}} \mathcal{B}_{1, \hat{\rho}}(\eta_1)\mathcal{B}_{2, \hat{\rho}}(\eta_2)\hat{\mathfrak{F}}_{\hat{\rho}}(\chi, \chi_R, \psi, \eta)\right].
\end{eqnarray}
where $\eta=\eta_1-\eta_2= \psi-\phi_R$ and the sum is over defect superconformal primaries $\hat \rho$ appearing in the bulk-defect OPE. 
To compute these, it is more convenient to further decompose the superconformal blocks with regard to the residual bosonic symmetry, with the blocks computed again as a product of \eqref{eq:CB-so13}, \eqref{eq:CB-so4R}, and \eqref{eq:CB-so2} 
\begin{equation}
    \hat{\mathfrak{F}}_{\hat{\rho
    }}= \sum_{\hat{\Delta}, n } C^{\hat{\rho}}_{\hat{\Delta}, n}\,(\eta) \, \mathfrak{f}_{\hat{\Delta}}\left(\chi\right)\,\mathfrak{f}^R_{[\frac{n}{2},\frac{n}{2}]}\left(\chi_R\right)\,  \mathfrak{h}_{\ell}(\psi).
\end{equation}
The decomposition is specified by the branching of $\hat \rho$ to the bosonic symmetry representations. 
Once we figure out the decomposition, we can fix the correct linear combination using the superconformal Ward identities. 
Note that the components of the supermultiplet all have the same charge under $SO(2)_t$. 

One of the simplest nontrivial examples is the symmetric representation with zero central charge $\hat{\mathcal{S}}_{0} = \expval{\mathcal V(1), 0}_{\mathrm{II} \pm} \otimes \expval{\mathcal V(1), 0}_{\mathrm{II} \pm}$. 
The spinless bosonic components with nonzero contribution to the bulk-defect OPE of a chiral primary operator are 
\begin{align}
    \hat{\mathcal{S}}_{0} \supset \left\{[\mathcal V(1), 0]\otimes [\mathcal V(1),0], \left [\mathcal V\left(\tfrac{1}{2}\right), \tfrac{1}{2}\right]\otimes \left[\mathcal V\left(\tfrac{1}{2}\right),\tfrac{1}{2}\right]\right\}
\end{align}
where $\mathcal V(\ell)$ are infinite-dimensional representations with either highest weight or lowest weight $\ell$, depending on the sign of $\ell$. 
The scaling dimensions of the bosonic states are $\hat \Delta = 1$ and $2$, and the $SO(4)_R$ representations are correspondingly $[\frac{1}{2}, \frac{1}{2}]$ and $[0, 0]$. 
The functions $C^{\hat{\rho}}_{\hat{\Delta}, n}(\eta)$ take into account the broken $SO(2)_R$ symmetry of each state in the multiplet. 
So, the most general ansatz for the superconformal block of $\hat {\mathcal{S}}_0$ is
\begin{equation}\label{eq:SCB-S0}
    \hat{\mathfrak{F}}_{\hat {\mathcal{S}}_{0}}= \sum_{k=0,1}\, c_k \, e^{\pm ik \eta} \, \mathfrak{f}_{k+1}(\chi)\, \mathfrak{f}^R_{[\frac{1-k}{2},\frac{1-k}{2}]}(\chi_R) .
\end{equation}
The coefficients $c_k$ are determined up to the overall scale by requiring this superconformal block to satisfy the superconformal Ward identities~\eqref{eq:SCWI-hol} and \eqref{eq:SCWI-antihol}. 
Remember that the dependence on the cross-ratios can be translated to that on the eigenvalues of the invariant quantity~\eqref{eq:x12-evals}.
The $\eta$ dependence is automatically ignored in the $v \rightarrow u$ and $\bar v \rightarrow u$ limits, which are translated to be $\eta \rightarrow 0$. 
Regarding these, the SCWIs fix the coefficients to be $c_0 = -c_1/2$.

We can also consider a slightly less simple example with the symmetric representations with some nonzero central charge $\ell$: $\hat{\mathcal{S}}_{ \ell} = \expval{\mathcal V(\frac{\ell}{2}), 0}_{\mathrm{I} -} \otimes \expval{\mathcal V(\frac{\ell}{2}), 0}_{\mathrm{I} -}$. 
The only states in these multiplets that can contribute to the defect OPE are the following spinless bosonic components 
\begin{equation}
 \hat {\mathcal{S}}_{\ell}\supset \{[\mathcal V(\tfrac{\ell}{2}), 0]\otimes [\mathcal V(\tfrac{\ell}{2}),0], [\mathcal V(\tfrac{\ell\pm 1}{2}), 1]\otimes[\mathcal V(\tfrac{\ell\pm 1}{2}),1], [\mathcal V(\tfrac{\ell\pm 2}{2}),0]\otimes[\mathcal V(\tfrac{\ell\pm 2}{2}),0]\},
\end{equation}
The plus signs are for $\ell \in \mathbb Z_+$ of lowest weight representations, and the minus signs are for $\ell \in \mathbb Z_-$ of highest weight representations. 
The scaling dimensions are $\hat{\Delta}= |\ell+1|+1, |\ell|+1, |\ell-1|+1$, and the corresponding $SO(4)_R$ representations are $[0,0], [\frac{1}{2},\frac{1}{2}], [0,0]$. 
The most general ansatz for the superconformal block of $\hat {\mathcal{S}}_\ell$ is
\begin{equation}\label{eq:SCB-Sl}
    \hat{\mathfrak{F}}_{\hat {\mathcal{S}}_{\ell}}= \sum_{k=-1,0,1}\, c_k \,e^{i \ell \psi} e^{ik \eta} \, \mathfrak{f}_{|\ell+k|+1}(\chi)\, \mathfrak{f}^R_{[\frac{1-|k|}{2},\frac{1-|k|}{2}]}(\chi_R) .
\end{equation}
The SCWIs \eqref{eq:SCWI-hol} and \eqref{eq:SCWI-antihol} fix the coefficients as $c_{-1}=c_{1}$ and $c_0=-c_{\pm}$.

\section{One-Point Functions: Non-protected operators} \label{one-point functions}
Our previous discussions focused on the correlation functions of the protected operators with a surface defect. 
We can more generally consider correlation functions involving non-protected single trace operators. 
Superconformal symmetry fixes the structure of the one-point function up to a function of the $\eta$, which we can assume to be a trigonometric polynomial from symmetry considerations:
\begin{equation}
    \langle \mathcal{O}_{\Delta}(P)\rangle= \frac{\mathcal{C}(\eta)}{(P\circ P)^{\Delta/2}}.
\end{equation}
At one-loop, scalar operators in the $SO(6)$ sector only mix among themselves and the mixing problem can be mapped to the diagonalization of the integrable $SO(6)$
Heisenberg chain. The R-matrix of the model is 
\begin{equation}
    R(u)= u(u+2) \mathbb{I}+ (u+2)\mathbb{P} -u \mathbb{K}.
\end{equation}
where the R-matrix and the operators are intertwiners of two $\mathbf 6$ representations corresponding to two neighboring spin sites. 
Specifically, $\mathbb I= \delta_I^{I'} \delta_J^{J'}$, $\mathbb P = \delta_I^{J'} \delta_J^{I'}$, and $\mathbb K = \delta_{IJ} \delta^{I'J'}$ in the matrix representation. 
The eigenstates of the Hamiltonian are labelled by three sets of Bethe roots $\{u_i, v_j, w_k\}$ satisfying a set of Bethe equations.
We denote the eigenstates as $\ket{\boldsymbol{u}}$. 
The leading-order result for the one-point function of the operator associated to the eigenstate $\ket{\boldsymbol{u}}$ is given by evaluating the operator on the classical values of the fields, which is equivalent to taking the overlap of $\ket{\boldsymbol{u}}$ with a matrix product state
\begin{equation}
\begin{aligned}
  \mathcal{C}(\eta)&= \bra{\textsf{MPS}(\eta)} \ket{\boldsymbol{u}}\\
  \ket{\textsf{MPS}(\eta)}&= \sum_{i} \tr[\omega_{I_1}\dots \omega_{I_L}]\ket{I_{1}\dots I_{L}}, \quad 
  \omega_{I_n} = \expval{\phi_{I_n}}.
\end{aligned}
\end{equation}
In order to avoid confusion with the spectral parameter dependence, we choose to omit the dependence on $\eta$ for the moment. 
For the case of a generic Gukov-Witten defect, the matrices $\omega_I$ are reducible in the sense that they are block diagonal whose structure depends on the breaking of the gauge group to the Levi subgroup $\mathbb L$, and each block behaves like a one-site product state.
We will comment on the singular cases later. 
In general, a state $\ket{\textsf{B}}$ is said to be integrable if it is annihilated by infinitely many charges of the system. 
More precisely, the state should satisfy~\cite{Piroli:2017sei}
\begin{equation}
    t(u)\ket{\mathsf{B}}=\Pi  t(u)\Pi \ket{\mathsf{B}},
\end{equation}
where $\Pi$ is the parity operator on the spin chain. 
This condition can be interpreted as the channel rotation of the boundary Yang-Baxter equation. 
It is known that for the case of a matrix product state, the integrability condition is satisfied if there exists a solution to the so-called \textit{square root relation} \cite{Pozsgay_2019}:
\begin{equation}
    \check{R}_{12}(u) \left(\psi(u)\otimes \omega\right)= \check{R}_{12}(u)\left(\omega \otimes \psi(u)\right),
\end{equation}
where we group the matrices in the MPS into a $\text{End}(\mathbb{C}^N)$-valued vector $\omega\in \mathbb{C}^6 \otimes \text{End}(\mathbb{C}^N)$, the hatted R-matrices are given by $\check{R}(u)= \mathbb{P} R(u)$, and we introduce an additional collection of matrices $\psi_{ab}(u) \in \mathbb{C}^6 \otimes \mathbb{C}^6 \otimes \text{End}(\mathbb{C}^N) $. 
The product $\otimes$ in $\omega \, \otimes\,  \psi(u)$ takes the tensor product of the $\mathbb C^6$ indices with the matrix products between their $\text{End}(\mathbb{C}^N)$ values. 
A solution of the square-root relation can then be used to construct solutions to the \textit{twisted} boundary Yang-Baxter equation. The solutions to these equations with various kinds of symmetries were studied in~\cite{Pozsgay_2019}.
The relevant symmetry pair for the Gukov-Witten defect is $(\mathfrak{so}(6),\mathfrak{so}(2)\oplus \mathfrak{so}(4))$. 
For one-site product states, the solution is:
\begin{equation}
    \psi_{ab}(u)= (2u+2)\,\omega_a \omega_b -u \, \omega_c \omega_d \,\delta^{cd}\,\delta_{ab}.
\end{equation}
Overlaps for this class of states were studied in \cite{DeLeeuw:2019ohp} and were used to compute one-point functions in the presence of BPS 't-Hooft loop operators \cite{Kristjansen:2023ysz, Kristjansen:2024map} and in the Coulomb branch \cite{Ivanovskiy:2024vel}; for a general formalism see \cite{Gombor:2024iix}. 

\subsection{$SL(2)$ Sector}
As a concrete example, we will take a particular $R$-symmetry polarization and study the leading-order one-point functions in that case.  
The simplest primary operators we can consider belong to the $SU(2)$ and $SL(2)$ closed sectors. The tree-level one-point functions for excited states in the $SU(2)$ sectors must vanish due to the $SO(4)_R$ symmetry of the defect regardless of how we orient the vacuum state inside of $SO(6)$.
One can understand it from the spin chain viewpoint as because the excited states belong to $SU(2)$ representations with one or more antisymmetric indices, and the classical field configurations commute. 
For the $SL(2)$ states, we can choose the operator to be 
\begin{equation}
\begin{aligned}
     O_{\Delta, S}(P)&= \sum_{n_1+\dots+n_L=S} \Psi_{n_1,\dots n_L}\tr[D^{n_1} \tilde{Z}\dots D^{n_L} \tilde{Z}]  \\
     D&= \partial_t- D_{x_1}\\
     \tilde{Z}&= \frac{1}{2}\left(Z+\bar{Z}+i(Y-\bar{Y})\right)\\
 z&= x_1 +i x_2
\end{aligned}
\end{equation}
In this case, the boundary state is similar to the one studied in \cite{Kristjansen:2023ysz} for `t Hooft loops \cite{Gombor:2021uxz}; it is given by
\begin{equation}
\begin{aligned}
 \ket{\mathsf{Bst}}&=  \tr[\ket{\mathsf{B}}^{\otimes L} ]\\
      \bra{\mathsf{B}}&= \sum_{n=0}^\infty \bra{0}\frac{(-K)^n}{n!} \big[\beta \cos((n+1)\psi) + \gamma \sin((n+1)\psi) \big]  \\    &=\sum_{n=0}^{\infty} (-1)^n (\beta^2+\gamma^2) T_{n+1} \left[\cos(\psi-\delta)\right]\bra{n}. 
\end{aligned}
\end{equation}
The only difference between this boundary state and the `t Hooft loop boundary state is that the Legendre polynomials are replaced by Chebyshev polynomials.
The overlap with a two magnon state can be found to be (up to an unimportant phase factor)
\begin{equation}
      \frac{\bra{\textsf{Bst}}\ket{\{u,-u\}
      }}{\bra{\{u,-u\}}\ket{\{u,-u\}}}= 2L \tr[\,(\beta^2+\gamma^2)(\beta \cos \psi+\gamma \sin \psi)^{L-2}\,] \sin^2\psi\,\frac{u}{\sqrt{L(L+1)(u^2+\frac{1}{4}})},
\end{equation}
which is the same as the two magnon overlap for the case of `t Hooft loop, up to a momentum independent polynomial in $\beta, \gamma$. 
We verified that for odd spin $S$ the overlap with the boundary state vanishes, but for even $S$ it does not satisfy the selection rules $\{\boldsymbol{u}\}=\{-\boldsymbol{u}\}$ which we take as evidence against integrability for the generic Gukov-Witten defect. 
This is somewhat similar to what occurs for three-point functions of non-protected operators with half-BPS sub-determinant operators. 
In that case, only a BPS special operator, the maximal giant graviton, leads to an integrable boundary state even though there is a large class of operators that preserve the same symmetry.   
Despite this, there are integrable subsectors corresponding to operators that do not carry the non-conserved charge in perturbation theory \cite{Berenstein:2006qk}. 
We expect something similar to occur for the surface defect case, meaning that at generic point of the moduli space, the defect breaks the integrability of the system for a generic sector due to the broken rotational symmetry. 
For the scalar $SO(6)$ sector, we expect that integrability is preserved at one-loop since we can construct a solution to the boundary Yang-Baxter equation, but mixing with operators outside this subsector should spoil this at higher loops.

\subsection{Rigid Limit: $SU(2)$ Sector}
Now we consider the singular case $\alpha=\beta=\gamma=0$. 
In that case, the field profiles are given by block diagonal matrices where each of the blocks satisfies the algebra of $SU(2)$. 
Without loss of generality, we can restrict to a single $k\times k$ block
\begin{equation}
\begin{aligned}
    Z^{cl}&= \frac{t_{1} - i t_2}{z \log r}\\
    A_\psi^{cl}&= \frac{t_3}{r \log r}.
\end{aligned}
\end{equation}
If we focus on operators belonging to the $SU(2)$ sector, the tree-level one-point functions are related to those of the D3-D5 defect set-up \cite{deLeeuw:2015hxa}, and the residual symmetry is related to the one relevant for studying three-point functions of half-BPS operators \cite{Basso:2015zoa, Jiang:2019xdz}. 
For example, we can restrict our attention to operators made out of two complex scalars
\begin{equation}
\begin{aligned}
        \tilde{Z}&= \frac{1}{2}\left(Z+\bar{Z}- i(Y-\bar{Y})\right)\\
        \tilde{Y}&=\frac{1}{2}\left(Y+\bar{Y}- i(Z-\bar{Z})\right),
\end{aligned}
\end{equation}
which correspond to the down and up spin states of the spin chain. The MPS in this case can be written as 
\begin{equation}
\begin{aligned}
  \ket{\textsf{MPS}_k}&= \sum_{i_n}\tr_k [M_{i_1}(\psi)\dots M_{i_L}(\psi)] \ket{i_1\dots i_L}\\
  M_{\downarrow}(\psi)&= t_1 \cos\psi - t_2 \sin \psi\\
   M_{\uparrow}(\psi)&= t_1 \sin\psi + t_2 \cos \psi.
\end{aligned}
\end{equation}
The matrices $M_{i}(\psi)$ are a $U(1)$ rotation of the standard basis for the $SU(2)$ algebra so they can be replaced by $t_1$ and $t_2$ inside the trace.
The result of the overlaps are the same as those found in \cite{deLeeuw:2015hxa, Buhl-Mortensen:2015gfd}, see also \cite{Jiang:2019xdz}. They are given by 
\begin{align}
\frac{\bra{\textsf{MPS}_k}\ket{\boldsymbol{u}}}{\bra{\boldsymbol{u}}\ket{\boldsymbol{u}}}&= 
2^{L-1} C_2\left(\left\{u_j\right\}\right) 
\sum_{j=\frac{1-k}{2}}^{\frac{k-1}{2}}  j^L \prod_{i=1}^{\frac{M}{2}} 
\,\frac{u_i^2\left(u_i^2 + \frac{k^2}{4}\right)}{\left[u_i^2+(j-\frac{1}{2})^2\right]
\left[u_i^2+(j+\frac{1}{2})^2\right]}\, \\
 C_2 \left(\left\{u_j\right\}\right) &=2\left[
 \left(\frac{2\pi ^2}{\lambda }\right)^L\frac{1}{L}
 \prod_{j}^{}\frac{u_j^2+\frac{1}{4}}{u_j^2}\,\,\frac{\det G^{SU(2)}_+}{\det G_-^{SU(2)}}\right]^{\frac{1}{2}}\\
\end{align}
where the Gaudin matrices $G_\pm^{SU(2)}$ are $M/2\times M/2$ given by 
\begin{equation}
\begin{aligned}
   \left( G^{SU(2)}_\pm\right)_{ij}&= \left[L \partial_u p(u_i)+\sum_{k=1}^{\frac{M}{2}}\mathcal{K}_+(u_i,u_k)\right]\delta_{ij}- \mathcal{K}_\pm(u_i,u_j)\\
   \mathcal{K}_\pm(u,v)&= \frac{1}{i}\left[\partial_u\log S^{SU(2)}(u,v)\pm \partial_u\log S^{SU(2)}(u,-v)\right].
\end{aligned}
\end{equation}
We expect that because the symmetry algebra for the rigid defects is an analytic continuation of that of the maximal giant graviton, the results of \cite{Jiang:2019xdz} can be applied to this case as well. We expect that the rigid defects remain integrable at finite 't Hooft coupling. This is somewhat clear in scalar sectors where the dilaton doesn't play a role. In sectors involving derivatives, one needs to be careful about using operators that are scale invariant when computing one-point functions as substituting the classical solutions at tree level leads to violations in the selection rules for the magnon momenta. These violations come from derivatives acting on the classical dilaton profile and they should be reabsorbed by using a Weyl invariant connection. This requires a more careful understanding of the quantization in the presence of the dilaton. We leave this for future work.

\section{Quantizing the Defect}\label{sec:propagators}
In this section, we perform the semiclassical quantization of $\mathcal N=4$ SYM in this surface defect background. This is a starting point for performing perturbative calculations. 
We expand the fields around their classical background values, and calculate the propagators of their quantum fluctuations. 
Similar calculations have been carried out in the domain wall and 't Hooft line backgrounds~\cite{Buhl-Mortensen:2016jqo, GimenezGrau:2018jyp, Gimenez-Grau:2019fld, Kristjansen:2023ysz}. The scalar propagators in the surface defect setting  appeared before in~\cite{Choi:2024ktc} and here we generalize by including fermionic fields. We also clarify the relation between field fluctuations and the defect primaries they couple to. 

First we are going to review some of the conventions for the $\mathcal{N}=4$ supersymmetry algebra in four dimensions. Usually it is easier to begin with $\mathcal{N}=1$ SUSY in ten-dimensions and to reduce it to 4d. The supercharges transform in one of the 16 dimensional spinor representation of $SO(1,9)$. The gamma matrices $\Gamma^M$, $M=0,1, \dots 9$ satisfy a Clifford algebra
\begin{equation}
    \{\Gamma^M, \Gamma^N\}= 2 g^{MN},
\end{equation}
and they can be thought of as either maps between the two spinor representations $\Gamma^I:\mathcal{S}^{\pm}\rightarrow \mathcal{S}^{\mp}$, or as a bilinear pairing $\Gamma^I: \mathcal{S}^{\pm}\times \mathcal{S}^{\pm}\rightarrow\mathbb{R}$ (in Lorentzian signature). The spinor representations are real and dual to each other in Lorentz signature, and they are distinguished by their eigenvalue under the chirality operator
\begin{equation}
    \bar{\Gamma}= \Gamma^0 \Gamma^1 \dots \Gamma^9,
\end{equation}
which acts as $\pm1$ on the $\mathcal{S}^\pm$. Another common notation are the gamma matrices with multiple indices $\Gamma^{I_1 I_2 \dots I_k}$ which is defined as the product $\Gamma^{I_1}\Gamma^{I_2}\dots \Gamma^{I_k}$ for pairwise distinct indices, and zero otherwise. This is equivalently the fully antisymmetrized product.
The lagrangian for 10d SYM takes the form
\begin{equation}
    S_{10d}= -\frac{1}{g^2}\int d^4x \Tr\left(\frac{1}{2}F^{MN}F_{MN}+i \bar{\lambda} \Gamma^I D_I \lambda\right).
\end{equation}
The supersymmetry generator in this case are constant spinors $\nabla_I \epsilon=0$ (i.e. Killing spinor) satisfying the chirality projection
\begin{equation}
\bar{\Gamma}\epsilon=\epsilon.
\end{equation}
The supersymmetry transformations are
\begin{equation}
\begin{aligned}
\delta A_I&= i \bar{\epsilon} \,\Gamma_I \lambda\\
\delta \lambda&= \Gamma^{IJ} F_{IJ} \epsilon.
\end{aligned}
\end{equation}
To obtain $\mathcal{N}=4$ SYM we reduce on six of the ten flat dimensions which reduces the 10d Lorentz symmetry to the 4d Lorentz symmetry and the R-symmetry as 
\begin{equation}
    SO(1,9)\rightarrow SO(1,3)\times SO(6)_R.
\end{equation}
The spinors $\epsilon$ can then be decomposed as follows. The 10d chirality matrix factorizes into a Lorentz and R-symmetry components
\begin{equation}
    \bar{\Gamma}= \hat{\Gamma} \Gamma'
\end{equation}
The eigenvalues of these chirality matrices are $\pm i$, so the eigenvalues of $\hat{\Gamma}$ and $\Gamma'$ have to be opposites. The spin representations of $SO(1,3)\sim SU(2)\times SU(2)$ are $(1,2)$ and $(2,1)$, and the spin representations of $SO(6)_R\sim SU(4)_R$ are the fundamental $\boldsymbol{4}$ and its complex conjugate $\bar{\boldsymbol{4}}$. So the 4d supersymmetries transform as a spinor in 
\begin{equation}
    \mathcal{S}^+= (\boldsymbol{2}, \boldsymbol{1}, \bar{\boldsymbol{4}})\oplus (\boldsymbol{1},\boldsymbol{2} , \boldsymbol{4})
\end{equation}
We now introduce the generic surface defect in the $M=0,3$ directions in the four-dimensional space. 
Among the Higgs scalars emerging from $A^{5,6,...,10}$ after the dimensional reduction, let us choose $A^5$ and $A^6$ to be the ones acquiring singular configurations. 
To carry out the perturbative calculations in the nontrivial background due to the surface defect, it is convenient to expand the fields around their classical singular configurations as 
\begin{align}
    A^M = \textsf{A}^M + a^M
\end{align}
Let us write the flat 10d metric taking polar coordinates for the transverse spatial direction to the defect: 
\begin{equation}
    ds^2=-dt^2+ dx^2+ dr^2+r^2 d\psi^2 + \sum_{i = 1}^6 dy_i dy_i,
\end{equation}
In this coordinate basis, the background fields are
\begin{equation}
\begin{aligned}
    \textsf{A}_\psi 
    &= \alpha\\
    \textsf{A}_5&= \frac{\beta\cos \psi + \gamma \sin \psi}{r}\\
    \textsf{A}_6&= \frac{\gamma\cos \psi -\beta \sin \psi}{r}.
\end{aligned}
\end{equation}
Alternatively we could use complex coordinates
\begin{equation}
\begin{aligned}
    ds^2&= -dt^2+dx^2 + 2dz d\bar{z}+ \sum_{a=1}^4d y_a dy_a + 2 dw d\bar{w}\\
\end{aligned}
\end{equation}
and the background fields are
\begin{equation}
\begin{aligned}
     \textsf{A}_z&=\frac{1}{2i}\frac{\alpha}{ z}\\ 
    \textsf{A}_w&= \frac{\beta+ i \gamma}{\sqrt{2}z}\\
\end{aligned}
\end{equation}
The classical configuration of the 10d field strength is 
\begin{equation}
\begin{aligned}
    \mathsf{F}_{z\bar{z}}&= \alpha\, \delta(z,\bar{z})\\
    \mathsf{F}_{M i}&= \mathsf{D}_M \mathsf{A}_i\\
    \mathsf{F}_{ ij}&= [\mathsf{A}_i, \mathsf{A}_j]=0
\end{aligned}
\end{equation}
where $\textsf{D}_M$ is the gauge covariant derivative with respect to the classical configuration of the gauge field, $\textsf{D}_M = \del_M - i[\textsf{A}_M, \cdot] $. 
These are the source of the defect CFT operators.

\subsection{Gauge Fixing}
We choose the gauge
\begin{equation}
    \textsf{D}_M A^M=0,
\end{equation}
which can be implemented with the gauge fixing term 
\begin{equation}
    \mathcal{L}_{gf}= -\frac{2}{g^2} \tr\left(\textsf{D}_M \bar{c} \,D^M c\right)
\end{equation}
After expanding the fields into the background and fluctuation parts, the action in this gauge is given by
\begin{equation}
\begin{aligned}
    S_{gf+ \Phi}&= -\frac{1}{g^2}\int d^4x \tr\bigg[ a_M\left(-\eta^{MN} \textsf{D}^2  + 2i \textsf{F}^{MN}\right)a_N + i \bar{\lambda} \Gamma^M \textsf{D}_M \lambda + 2 \bar{c}(- \textsf{D}^2) c\\
    &- 2 i \textsf{D}_M a_N [a^M, a^N]+ \bar{\lambda} \Gamma^M[a_M, \lambda]+ 2i \textsf{D}_M \bar{c}\,[a^M, c]-\frac{1}{2}[a_M, a_N]^2\bigg].
\end{aligned}
\end{equation}
We can classify the fields into easy and hard fields depending on whether they mix or not due to their nonzero background values. 
The nontrivial mixing between the scalars and vector fields comes from $\mathsf{F}_{Mi}$ for $i=5,6$.
The second covariant derivative term is (in polar coordinates)
\begin{equation}
\begin{aligned}
\tr\left[a\,\textsf{D}^2 a\right]&= \tr\left[a\partial^2 a+2i a\,[\mathsf{A}^M, \partial_M a]-a\,[\mathsf{A}^M, [\mathsf{A}_M, a]]\right]\\
&=\tr\left[a\left(\eta^{\alpha \beta}\partial_{\alpha}\partial_{\beta}+ \partial_r^2 +\frac{1}{r}\partial_r \right)a + a \textsf{D}^\psi \textsf{D}_\psi a-a[\mathsf{A}^5, [\mathsf{A}_5, a]]-a[\mathsf{A}^6, [\mathsf{A}_6, a]]\right]
\end{aligned}
\end{equation}
The indices $\alpha, \beta = 0, 3$ run for the parallel directions to the defect.

\subsection{Propagators}
Now we can consider the quadratic action for an arbitrary Gukov-Witten defect. The first step is to get rid of the non-trivial derivatives in $\psi$. These terms only appear for $\alpha\neq0$ and they make the fields transform as a non-trivial line bundle on $S^1$. We can eliminate the $\alpha $ dependence by introducing the modified fields 
\begin{equation}
    a= \mathcal{U}_\alpha^\dagger \,\tilde{a}\, \mathcal{U}_\alpha= e^{-i \psi \alpha} \Tilde{a} \,e^{i \psi \alpha}.
\end{equation}
The action of the background covariant derivatives $\textsf{D}_z$ on $a$ results in regular derivatives acting on $\tilde{a}$. Substituting these fields into the action reduces it to the case with $\alpha=0$, so no local operator can detect non-zero $\alpha$ as expected. From now on we drop the tildes, after which the 10d Kinetic term common to all bosons except $a_{r, \psi}$ reduces to 
\begin{equation}
\begin{aligned}
\tr\left[a\,\textsf{D}^2 a\right]&=\tr\left[a\left(\eta^{\alpha \beta}\partial_{\alpha}\partial_{\beta}+ \partial_r^2 +\frac{1}{r}\partial_r +\frac{1}{r^2}\partial_\psi^2\right)a -\frac{1}{r^2}a[\beta, [\beta, a]]-\frac{1}{r^2}a[\gamma, [\gamma, a]]\right]
\end{aligned}
\end{equation}
To simplify the derivative terms we do a field redefinition $a=  a'/r$, and passing the derivatives through we get
\begin{equation}
\begin{aligned}
\tr\left[a\,\textsf{D}^2 a\right]&=r^{-4}\tr\left[a'\left(r^2\eta^{\alpha \beta}\partial_{\alpha}\partial_{\beta}+ r\partial_r^2 -r\partial_r +1+\partial_\psi^2\right)a' -a'[\beta, [\beta, a']]-a'[\gamma, [\gamma, a']]\right]\\
&=r^{-1}\sqrt{g_{AdS_3\times S^1}}\tr\left[a'\left(\Delta_{AdS_3\times S^1}+1\right)a' -a'[\beta, [\beta, a']]-a'[\gamma, [\gamma, a']]\right]
\end{aligned}
\end{equation}
The additional factor of $1/r$ cancels precisely with the volume factor of the flat space metric with polar coordinates along two of the directions. For $M=5,\dots, 10$ this is a Weyl transformation and the additional mass term arises from the conformal coupling to the background. For $a_{r, \psi}$ the kinetic terms take the form
\begin{equation}
\begin{aligned}
\tr\left[a_r\,\textsf{D}^2 a_r\right]&=\tr\bigg[a_r\left(\eta^{\alpha \beta}\partial_{\alpha}\partial_{\beta}+ \partial_r^2 +\frac{1}{r}\partial_r +\frac{1}{r^2}\partial_\psi^2-\frac{1}{r^2}\right)a_r-\frac{2}{r^2}a_r\partial_\psi a_\psi \\
&-\frac{1}{r^2}a_r[\beta, [\beta, a_r]]-\frac{1}{r^2}a_r[\gamma, [\gamma, a_r]]\bigg]\\
\tr\left[a_\psi\,\textsf{D}^2 a_\psi\right]&=\tr\bigg[a_\psi\left(\eta^{\alpha \beta}\partial_{\alpha}\partial_{\beta}+ \partial_r^2 +\frac{1}{r}\partial_r +\frac{1}{r^2}\partial_\psi^2-\frac{1}{r^2}\right)a_\psi +\frac{2}{r^2}a_\psi\partial_\psi a_r\\
&-\frac{1}{r^2}a_\psi[\beta, [\beta, a_\psi]]-\frac{1}{r^2}a_\psi[\gamma, [\gamma, a_\psi]]\bigg].
\end{aligned}
\end{equation}
After rescaling by $1/r$ the only modification is an additional $-1$ in the mass term and the cross term with the single $\psi$ derivative. 
Next we should deal with the coupling to the background field strength $\mathsf{F}^{MN}$. If we regularize the singularity by imposing a cut-off at $r=\epsilon$, we can ignore the coupling to $F^{z\bar{z}}$. 
First we work in complex coordinates where the terms take the form
\begin{equation}
\begin{aligned}
 \tr[a_M \mathsf{F}^{MN}\circ a_N]
 &=\frac{1}{\sqrt{2}r^2}\bigg(\tr\bigg[-a_r [(\beta-i\gamma)e^{i \psi},a_w]+a_w [(\beta-i\gamma)e^{i \psi}, a_r]\\
 &+i a_\psi[(\beta-i\gamma)e^{i \psi},a_w]
 -ia_w[(\beta-i\gamma)e^{i \psi},a_\psi]\bigg]+c.c.\bigg.
\end{aligned}
\end{equation}
Then we should eliminate the factors of $e^{i\psi}$; we can do this by turning on a background $U(1)_R$ gauge field
\begin{equation}
\begin{aligned}
    a_w&= e^{-i q_R\psi}\varphi_w= e^{-i \psi}\varphi_w\\
    \partial_\psi a_w &= (\partial_\psi- i q_R )\varphi_w.
\end{aligned}
\end{equation}
This is consistent with the supersymmetry of the theory on $AdS_3\times S^1$; after this we perform the rescaling $a= a'/r$ which gives
\begin{equation}
\begin{aligned}
 \tr[a_M \mathsf{F}^{MN}\circ a_N]
 &=\frac{1}{\sqrt{2}r^4}\bigg(\tr\bigg[-a_r' [(\beta-i\gamma),\varphi_w']+\varphi_w' [(\beta-i\gamma), a_r']\\
 &+i a_\psi'[(\beta-i\gamma),\varphi_w']
 -i\varphi_w'[(\beta-i\gamma),a_\psi']\bigg]+c.c.
\end{aligned}
\end{equation}
and all the bosons are now written covariantly in $AdS_3\times S^1$. The bosons naturally split into easy/transverse and hard/longitudinal modes. The transverse modes are $a_{0,3}$, $\phi_{a \dot{a}}= a_{7,8,9,10}$, and the ghosts $c$. The fields $\phi_I$ transform as bi-spinors of the residual $SO(4)_R= SU(2)\times SU(2)$ symmetry. The hard modes are the normal components of the gauge field to the boundary $a_{r,\psi}$ and the complex scalar $\varphi_w$. Both of these modes are vectors due to the one-form nature of $\phi= \varphi_w dz + \varphi_{\bar{w}} d\bar{z}$. As explained in \cite{Aharony:2010ay}, vector fields in $AdS$ can be described in terms of conformally coupled scalars, which is precisely what we have with our choice of gauge. The mass terms are only nontrivial for off-diagonal components of the fields. The common mass term for all the fields is
\begin{equation}
|\mathfrak{Z}_{ij}|^2= (\beta_i-\beta_j)^2+(\gamma_i-\gamma_j)^2,
\end{equation}
where $i,j$ denote the $U(N)$ matrix indices of the fields.
\subsection{Easy Bosons}
The propagator for the easy scalar is the solution to 
\begin{equation}
\left[-\Delta_{AdS_3}- \partial_\psi^2+ (|\mathfrak{Z}_{ij}|^2-1)\right] G(x,x', \psi, \psi')=0. 
\end{equation}
We can solve this equation by performing a KK-reduction on $S^1$ 
\begin{equation}
\begin{aligned}
a(\psi, x)&= \sum_{\ell=-\infty}^{\infty} e^{i \ell \psi} a^{(\ell)}(x),
\end{aligned}
\end{equation}
where the reality condition on the fields implies
\begin{equation}
    (a^*)^{(\ell)}= a^{(-\ell)},
\end{equation}
so we are left with a tower of conformally coupled scalars in $AdS_{3}$ with masses $\left(m^{\boldsymbol{s}}_{ij}\right)^2= \ell^2+|\mathfrak{Z}_{ij}|^2-1 $. These fields couple to scalar operators on the defect with conformal dimensions
\begin{equation}
    \hat{\Delta}^{\boldsymbol{s}}= 1 +\sqrt{\ell^2+ |\mathfrak{Z}_{ij}|^2}.
\end{equation}
The bulk-to-bulk propagator for each of the KK modes is given by 
\begin{equation}\label{eq:ads3-propagator}
    G_{\hat{\Delta}}(x_1, x_2)= \frac{\tilde{\chi}^{\Hat{\Delta}}}{2^{\hat{\Delta}+1}}\, _2F_1\left(\frac{\hat{\Delta}}{2}, \frac{\hat{\Delta}}{2}+\frac{1}{2}; \hat{\Delta}; \tilde{\chi}^2\right),
\end{equation}
where $\tilde{\chi}$ is the geodesic distance between the operator insertions which is given by 
\begin{align}
    \Tilde \chi = \frac{2r_1r_2}{r_1^2 + r_2^2 + (\Vec x_1 - \Vec x_2)^2}
\end{align}
For diagonal modes $i=j$, the fields couple to operators belonging to family of short representations of the defect superconformal algebra $\hat{\Delta}= 1+|\ell|$, the symmetric representations. Off-diagonal modes belong to short multiplets of a central extension of the defect superconformal algebra with $\mathfrak P$ and $\mathfrak K$, so we expect their dimensions to be protected from quantum corrections. These representations are related to the defect BPS by an outer automorphism \cite{Beisert:2006qh}.
\subsection{Hard Bosons}
The kinetic terms for the $\varphi_w$ modes are of the form
\begin{equation}
\begin{aligned}
2 \pi\sum_{\ell} \int d^3x \sqrt{g_{AdS_3}}\bigg(&(\varphi_{\bar{w}})_{ij}^{(-\ell)}\left(\Delta_{AdS_3}-(|\mathfrak{Z}_{ij}|^2+(\ell-1)^2-1)\right)(\varphi_w)_{ji}^{(\ell)} \\
  +&(\varphi_{w})_{ij}^{(-\ell)}\left(\Delta_{AdS_3}-(|\mathfrak{Z}_{ij}|^2+(\ell+1)^2-1)\right)(\varphi_{\bar{w}})_{ji}^{(\ell)}\bigg).
\end{aligned}
\end{equation}
The reality condition on the fields implies
\begin{equation}
\left(\varphi_w^{(\ell)}\right)^*= \varphi_{\bar{w}}^{(-\ell)}.
\end{equation}
For the hard bosons we first perform the KK reduction on $S^1$. Because of the background $U(1)_R$ gauge field the diagonal terms of the mass matrix for $\varphi_w$ are shifted relative to  those of $a_{r, \psi}$ . This gives a mass matrix of the following form for the fields in $AdS_3$:
\begin{equation}
    \begin{pmatrix}
        a_r^{(\ell)}\\
        a_{\psi}^{(\ell)}\\
        \varphi_w^{(\ell)}\\
        \varphi_{\bar{w}}^{(\ell)}
    \end{pmatrix}_{ij}^*\begin{pmatrix}
    (|\mathfrak{Z}_{ij}|^2+ \ell^2)&2 i \ell&-i\sqrt{2}\, \mathfrak{Z}^\dagger_{ij}&-i\sqrt{2}\, \mathfrak{Z}_{ij}\\
    -2i \ell&(|\mathfrak{Z}_{ij}|^2+ \ell^2)&-\sqrt{2}\, \mathfrak{Z}^\dagger_{ij}&\sqrt{2}\, \mathfrak{Z}_{ij}\\
    i\sqrt{2}\, \mathfrak{Z}_{ij}&-\sqrt{2}\, \mathfrak{Z}_{ij}&(|\mathfrak{Z}_{ij}|^2+ (\ell-1)^2-1)&0\\
    i\sqrt{2}\, \mathfrak{Z}^\dagger_{ij}&\sqrt{2}\, \mathfrak{Z}_{ij}^\dagger&0&(|\mathfrak{Z}_{ij}|^2+ (\ell+1)^2-1)
    \end{pmatrix} \begin{pmatrix}
        a_r^{(\ell)}\\
        a_{\psi}^{(\ell)}\\
        \varphi_w^{(\ell)}\\
        \varphi_{\bar{w}}^{(\ell)}
    \end{pmatrix}_{ji}. 
\end{equation}
This is slightly different from the conventions used in \cite{Choi:2024ktc} for their background field computation. 
The eigenvalues of the $AdS_3$ mass matrix are
\begin{equation}
    (m_{\pm}^{ij})^2= \left(\sqrt{|\mathfrak{Z}_{ij}|^2+\ell^2}\pm1\right)^2-1;
\end{equation}
note the $-1$ corresponding to the conformal coupling to the curvature. These correspond to 1-form operators on the boundary with conformal dimensions
\begin{equation}
\hat{\Delta}_{\pm}^{\boldsymbol{v}}= \sqrt{|\mathfrak{Z}_{ij}|^2+\ell^2}\pm1.
\end{equation}
\subsection{Fermions}
For the fermion fields we need to reduce the spinors from 10d to 3d in two steps; by first reducing to $\mathbb{R}^{1,3}$ and then doing a further $S^1$ reduction; we also have to decompose the $SO(6)_R$ labels. Our conventions for the gamma matrices are such that $\Gamma^0$ in the Lorentz signature is related to $\Gamma^4$ in Euclidean signature by multiplication by $i$, and so the indices run from $\mu=0,1,2,3$ and $I=5,6,7,8,9,10$. The gamma matrices decompose as follows
\begin{equation}
\begin{aligned}
    \Gamma^\mu&= \gamma^\mu \otimes \mathbf{1}_{2}\otimes\mathbf{1}_4\;\;\;\; \mu=0,1,2,3\\
    \Gamma^{5,6}&=\gamma_5\otimes \sigma^{1,2}\otimes \mathbf{1}_4\\
    \Gamma^{A}&=\gamma_5\otimes \sigma^{3}\otimes \tilde{\gamma}^{A-6},\;\;\;\; A=7,8,9,10
\end{aligned}
\end{equation}
where $\gamma^\mu$ and $\Tilde \gamma^{A}$ are $SO(1,3)$ and $SO(4)_R$ gamma matrices. The $SO(1,9)$ quantum numbers split into $SO(1,3)\times SO(2)_R\times SO(4)_R$ labels. It will be convenient to split the 10d chiral spinor into eigenstates of the chirality matrices for each factor;
\begin{equation}
\begin{aligned}
\gamma_5&=i \gamma^0 \gamma^1 \gamma^2\gamma^3\\
\tilde{\gamma}^5&= -\tilde{\gamma}^1 \tilde{\gamma}^2\tilde{\gamma}^3\tilde{\gamma}^4\\
  \Gamma^{11}&= (-i\gamma^5)\otimes (i \sigma^3)\otimes \tilde{\gamma}^5
\end{aligned}
\end{equation}
The $SO(2)_R$ charge generator will also be important 
\begin{equation}
    \frac{i}{4}[\Gamma^{5}, \Gamma^6]= \mathbf{1}_4\otimes\left(-\frac{1}{2}\sigma^3\right)\otimes \mathbf{1}_4.
\end{equation}
In this convention, the $SO(2)_R$ charge carries the opposite sign relative to the 2d chirality.
Then a chiral spinor in 10d decomposes into 
\begin{equation}\label{eq:10d-spinor-decomposition}
\begin{aligned}
      \mathbf{16}_+&\rightarrow \left[(1/2,0)\otimes(-1/2)\otimes(1/2, 0)\right]\oplus \left[(0,1/2)\otimes(+1/2)\otimes(1/2,0)\right]\\
      &\oplus \left[(1/2,0)\otimes(+1/2)\otimes(0,1/2) \right]\oplus \left[(0,1/2)\otimes(-1/2)\otimes(0,1/2)\right],
\end{aligned}
\end{equation}
and so we can think of a 10d chiral spinor as a pair of 4d Dirac fermions tranforming in the $(1/2,0)$ and $(0,1/2)$ representation of the residual $SO(4)_R$ symmetry. Next we need to impose the 10d Majorana condition. The reality condition on the gamma matrices in Lorentz signature is
\begin{equation}
    \Gamma^*= \mathcal{C}\Gamma^0 \Gamma \left(\mathcal{C}\Gamma^0\right)^{-1}
\end{equation}The charge conjugation matrix $\mathcal{C}=\Gamma^{1}\Gamma^3 \Gamma^6 \Gamma^7 \Gamma^9 $decomposes as
\begin{equation}
    \mathcal{C}= \gamma^{1}\gamma^3 \gamma^5 \otimes \sigma^2 \otimes \tilde{\gamma}^1 \tilde{\gamma}^3,
\end{equation}
where the first factor can be taken to be the 4d charge conjugation matrix, the second factor is also the charge conjugation in 2d while the last factor is the time reversal matrix in 4d. The Majorana condition for a 10d spinor $\lambda$ is then 
\begin{equation}
    \lambda^*= \mathcal{C} \Gamma^0 \lambda= (\mathcal{C}_4 \gamma^0 \otimes \mathcal{C}_2 \otimes \tilde{\mathcal{T}}_4) \lambda,
\end{equation}
and since we want to reduce to 4d it will be useful to relate this condition to the Majorana condition in 4d. To understand this, we can decompose the 10d spinor into its components under~\eqref{eq:10d-spinor-decomposition}; 
\begin{equation}
\begin{aligned}
    \lambda&= \chi_a\otimes \epsilon^-\otimes \hat{c}^{a}+ \eta^\dagger_a \otimes \epsilon^+ \otimes \hat{c}^{a}\\
    &+ \xi_{\dot{a}}\otimes \epsilon^+ \otimes \hat{c}^{\dot{a}}+ \zeta^\dagger_{\dot{a}} \otimes \epsilon^-\otimes \hat{c}^{\dot{a}}
\end{aligned}
\end{equation}
where we suppressed the Lorentz spinor indices. Clearly the Majorana condition does not mix the dotted and undotted $SO(4)_R$ indices which suggest that we should pair $\chi$ and $\eta^\dagger$ into a Dirac spinor and similarly for the remaining spinors. For a moment, let us ignore the $SO(4)$ indices; we can always multiply $\mathcal{C}$ with a phase such that $\mathcal{C}_2$ acts as follows on $\epsilon^{\pm}$
\begin{equation}
    \mathcal{C}_2 \epsilon^{\pm}= \epsilon_{\mp}
\end{equation}
which we should understand as maps between the two different $SO(2)$ chiralities. Complex conjugation raises and lowers the indices of $\epsilon^{\pm}$ but does not mix chiralities, and $\mathcal{C}_2$ maps the chiralities into each other while raising/lowering the indices. So we can define component-wise
\begin{equation}
    \epsilon^+= -\epsilon_-
\end{equation}
and so on. Similarly $\mathcal{C}_4\gamma^0$ changes the Lorentz chirality without lowering/raising the spinor indices:
\begin{equation}
    \left(\mathcal{C}_4\gamma^0\otimes \mathcal{C}_2\right)\begin{pmatrix}
        \chi_{\alpha} \otimes \epsilon^{-}\\
        \eta^{\dagger \dot{\alpha}} \otimes \epsilon^+
    \end{pmatrix}= \begin{pmatrix}
        \eta^{\dagger\dot{\beta}}\otimes \epsilon_-\\
        \chi_{\beta}\otimes \epsilon_+
    \end{pmatrix}.
\end{equation}
For the last factor we can use $\tilde{\mathcal{T}}_4$ to raise and lower the remaining indices. Complex conjugation raises and lowers the indices as well without exchanging the chiralities. Putting everything together, this implies that
\begin{equation}
\begin{aligned}
        \chi&= \eta\\
        \xi&= \zeta,
\end{aligned}
\end{equation}
which means that we have four Majorana fermions in 4d after dimensional reduction. 
The kinetic term is
\begin{equation}
\begin{aligned}
\tr\left[  \bar{\lambda} (\gamma^\mu \partial_\mu \otimes \mathbf{1}_2\otimes \mathbf{1}_4) \lambda +\frac{1}{ \bar{z}} \bar{\lambda} (\gamma^5\otimes\sigma^+\otimes \mathbf{1}_4) [\mathfrak Z^\dagger,\lambda]+ \frac{1}{z}\bar{\lambda} (\gamma^5\otimes\sigma^-\otimes \mathbf{1}_4)[\mathfrak Z,\lambda]\right].
\end{aligned}
\end{equation}
Since this expression is $SO(4)_R= SU(2)\times SU(2)$ symmetric we can ignore the $SO(4)_R$ indices for the time being. To simplify the dependence on the coordinates we should turn on a background $SO(2)$ field
\begin{equation}
    \lambda= e^{\frac{i}{2}( \mathbf{1}_4\otimes\sigma^3\otimes \mathbf{1}_4) \psi}\,\lambda_0;
\end{equation}
the reason we do this is so that the $\psi$ dependence on the mass term cancels
\begin{equation}
    \frac{1}{\bar{z}} \bar{\lambda} (\gamma^5\otimes\sigma^+\otimes \mathbf{1}_4) [\mathfrak{Z}^\dagger,\lambda ]=  \frac{1}{r} \bar{\lambda}_0 (\gamma^5\otimes\sigma^+\otimes \mathbf{1}_4) [\mathfrak{Z}^\dagger,\lambda_0 ]
\end{equation}
From now on we will drop the lower label and use $\lambda$ to represent the spinor after including the background gauge field. Defining the 4d Majorana spinors 
\begin{equation}
    \lambda^+= \begin{pmatrix}
        \chi\\ \chi^\dagger
    \end{pmatrix}, \;\;\;\lambda^-= \begin{pmatrix}
        \xi\\ \xi^\dagger
    \end{pmatrix},
\end{equation}
the mass terms can be rewritten in the form
\begin{eqnarray}
    \frac{1}{r}\tr\left[\xi_{\dot{a}} [\mathfrak{Z}, \xi^{\dot{a}}]+ \xi^\dagger_{\dot{a}} [\mathfrak{Z}^\dagger, \xi^{\dagger\dot{a}}]+ \chi_a [\mathfrak{Z}^\dagger, \chi^a]+\chi^\dagger_{a} [\mathfrak{Z}, \chi^{\dagger a}]\right]
\end{eqnarray}

\subsubsection{Mapping to $AdS_3$}
Now we can perform the boundary expansion of the fields by performing a field redefinition
\begin{equation}
    \lambda\rightarrow\frac{\lambda'}{r^{3/2}}
\end{equation}
after which the fermion kinetic terms become
\begin{equation}
    -\frac{1}{g^2}\int \sqrt{g_{AdS_3\times S^1}} d^4x \,\tr\left[ i\bar{\lambda}' \left( e^{\mu}_{\hat{\alpha}}\,\Gamma^{\hat{\alpha}} \partial_{\mu}-\Gamma^{r}+\frac{i}{2} \Gamma^{\psi}\left(\mathbf{1}_4\otimes \sigma^3 \otimes \mathbf{1}_4\right)\right)\lambda'\right]
\end{equation}
This is the action of a fermion on $AdS_3\times S^1$ with a background gauge field along the circle direction. The action of $\mathbf{1}_4\otimes\sigma^3 \otimes \mathbf{1}_4$ can be replaced with $ \pm\gamma_5  $ on $\lambda^\pm$, with the upper indices denoting the $SO(4)_R$ chirality, due to the 10d chirality of the spinor which allows us to rewrite everything in 4d language:
\begin{equation}
\begin{aligned}
    &-\frac{1}{g^2}\int \sqrt{g_{AdS_3\times S^1}} d^4x \,\tr\left[ i\bar{\lambda}^{+\,a} \left( e^{\mu}_{\hat{\alpha}}\,\gamma^{\hat{\alpha}} \nabla_\mu+\frac{i}{2} \gamma^{\psi}\gamma^5\right)\lambda^{+}_a + i\bar{\lambda}^{-\,\dot{a}} \left( e^{\mu}_{\hat{\alpha}}\,\gamma^{\hat{\alpha}} \nabla_\mu-\frac{i}{2} \gamma^{\psi}\gamma^5\right)\lambda^{-}_{\dot{a}}\right]  \\
    &=-\frac{1}{g^2}\int \sqrt{g_{AdS_3\times S^1}} d^4x \,\tr\left[ i\bar{\lambda}^{+\,a} \left( e^{\mu}_{\hat{\alpha}}\,\gamma^{\hat{\alpha}} \nabla_\mu+\frac{1}{2} \gamma^{\star}\right)\lambda^{+}_a + i\bar{\lambda}^{-\,\dot{a}} \left( e^{\mu}_{\hat{\alpha}}\,\gamma^{\hat{\alpha}} \nabla_\mu-\frac{1}{2} \gamma^*\right)\lambda^{-}_{\dot{a}}\right],
\end{aligned}
\end{equation}
where $\gamma^*= \gamma^0 \gamma^{1}\gamma^{2}$ is the product over the gamma matrices in the $AdS$ directions. To reduce from 4d to 3d we should pick a basis of gamma matrices that remains real after the dimensional reduction such that the 3d spinors are Majorana. In such a basis the 4d chirality matrix is not diagonal, so the mass term is not as simple. For example we can take the 4d gamma matrices in a basis 
\begin{equation}
\begin{aligned}
    \gamma^{0,1,2}&= \begin{pmatrix}
        0& \hat{\gamma}^{0,1,2}\\
        \hat{\gamma}^{0,1,2}&0
    \end{pmatrix},\;
    \;
    \gamma^3= \begin{pmatrix}
        \mathbf{1}_2 &0 \\
        0& -\mathbf{1}_2
    \end{pmatrix}\\
     \gamma^5&= \begin{pmatrix}
                0& -i \mathbf{1}_2\\
                i \mathbf{1}_2&0
            \end{pmatrix}.
\end{aligned}
\end{equation}
where $\hat \gamma^k$ are the 3d gamma matrices. 
In this basis $\gamma^0 \mathcal{C}_4$ is equal to the identity. 
Since $\tilde{\mathcal{T}}_4$ commutes with the $SO(4)$ chirality matrix, we can formally diagonalize $\mathcal{C}_2\otimes\tilde{\mathcal{T}}_4$\footnote{We say formally because this matrix is a map between complex conjugate representations.} and restrict to the subspace with eigenvalue equal to one. The eigenvalues of $\mathcal{C}_2$ and $\tilde{\mathcal{T}}_4$ are $\pm i$ 
 if we work with a convention such that $\mathcal{C}^T= \mathcal{C}^{-1}$. In this basis the 10d Majorana condition is just
\begin{equation}
    \lambda^*= \lambda,
\end{equation}
and the 10d spinor splits into four Majorana spinors
\begin{equation}
    \Psi_{+}^{-}, \;\;\;\Psi_{-}^{+},\;\;\;\tilde{\Psi}_{\dot{+}}^{-},\;\;\;\tilde{\Psi}_{\dot{-}}^{+},
\end{equation}
where now the upper indices denote the sign of the eigenvalue of $\mathcal{C}_2$, and the lower index labels the $SU(2)\times SU(2)$ basis where $\mathcal{T}_4$ is diagonal. Because the signs are always opposite the upper indices are redundant and we will omit them. To reduce from 4d to 3d we simply split the four components into pairs of two-dimensional real spinors which are eigenvectors of $\gamma^\psi =\gamma^3$;
\begin{equation}
    \Psi_{\pm}= \begin{pmatrix}
        \chi_{\pm}\\
        \eta_{\pm}
    \end{pmatrix},
\end{equation}
and similarly for the fields with negative $SO(4)_R$ chirality.
The 10d chirality condition allows us to replace the background $SO(2)$ gauge field by a chirality transformation
\begin{eqnarray}
    \Gamma^{11}\lambda=\lambda\Rightarrow \left(\mathbf{1}_4\otimes \sigma^3 \otimes \mathbf{1}_4\right)\lambda= 
   \left( \gamma^5\otimes \mathbf{1}_2\otimes\tilde{\gamma}^5 \right)\lambda.
\end{eqnarray}
so we just need to split kinetic terms for the fermions according to their $SO(4)_R$ chirality. The kinetic terms (for one of the $SO(4)_R$ chiralities) are then
\begin{equation}
\begin{aligned}
    &=\tr\left[ i\bar{\Psi}^{\,a} \left( e^{\mu}_{\hat{\alpha}}\,\gamma^{\hat{\alpha}} \nabla_\mu+\frac{1}{2} \gamma^{\star}\right)\Psi_a \right]\\
    &=\tr\left[ i\bar{\chi}^{\,a} \left( e^{\mu}_{\hat{\alpha}}\,\hat{\gamma}^{\hat{\alpha}} \nabla_\mu+\frac{1}{2}\right)\chi_a +i\bar{\eta}^{\,a} \left( e^{\mu}_{\hat{\alpha}}\,\hat{\gamma}^{\hat{\alpha}} \nabla_\mu +\frac{1}{2}\right)\eta_a + i \bar{\eta}^a\partial_\psi\chi_a -  i \bar{\chi}^a\partial_\psi \eta_a\right].
\end{aligned}
\end{equation}
An important comment is that the barred spinors in the second line are defined in 3d, $\bar\chi= \chi^\dagger \hat{\gamma}^0$, whereas the barred spinors in the first line are 4d spinors. Now can easily perform the $S^1$ reduction 
\begin{eqnarray}
    \chi(\psi)= \sum_\ell e^{i \ell \psi} \chi^{(\ell)}
\end{eqnarray}
and similarly for all the other modes.
To simplify the mass terms we need to compute the matrix elements of $\gamma^5\otimes \sigma^\pm \otimes \mathbf{1}$ in the Majorana basis. In the basis where $\sigma^3$ is diagonal, the eigenvectors of $\mathcal{C}_2\otimes \mathcal{T}_4$ are given by
\begin{equation}
  \frac{1}{\sqrt{2}} \Psi_{\mp} \otimes\begin{pmatrix}
        1\\
        \pm i
    \end{pmatrix}= \Psi_{\mp}\otimes v_{\pm}.
\end{equation}
Here we should be careful when taking the complex conjugate, since the basis vectors themselves are complex but the coefficients of the spinors are real. After assembling the vectors $v_\pm$ into a matrix $V$, the relevant matrix elements are
\begin{equation}
\begin{aligned}
M^a_{b}= V^\dagger\sigma^+ V&=\frac{1}{2}\begin{pmatrix}
       1&-i\\
       -i&-1
    \end{pmatrix}\\
  M^{\dagger\, a}_{b}=  V^\dagger\sigma^- V&=\frac{1}{2}\begin{pmatrix}
       1&i\\
       i&-1
    \end{pmatrix}
\end{aligned}
\end{equation}
All the terms are $SO(2,2)$ invariant so the mass terms are diagonal in the $AdS_3$ spinor space; the resulting mass matrix for the positive $SO(4)_R$ chirality modes is 

\begin{equation}
    \begin{pmatrix}
        \bar{\chi}^a\\
        \bar{\eta}^a
    \end{pmatrix}\begin{pmatrix}
        \mathfrak{Z} M^{\dagger\,b}_a + \mathfrak{Z}^\dagger M_a^b + \frac{1}{2}\delta_a^b& i \ell \delta_a^b\\
        -i \ell\delta_a^b& -\mathfrak{Z}  M{^\dagger\,b}_a  - \mathfrak{Z}^\dagger M_a^b + \frac{1}{2}\delta_a^b
    \end{pmatrix}\begin{pmatrix}
        \chi_b\\
        \eta_b
    \end{pmatrix}.
\end{equation}
The eigenvalues are doubly degenerate and given by:
\begin{equation}
    \left(m^+_{f,\pm}\right)_{ij}\in \left\{\frac{1}{2}+ \sqrt{\ell^2 + |\mathfrak{Z}_{ij}|^2},-\left(-\frac{1}{2}+\sqrt{\ell^2 + |\mathfrak{Z}_{ij}|^2} \right)\right\}.
\end{equation}
From this we can infer the eigenvalues for the remaining modes (negative $SO(4)_R$ chirality)
\begin{equation}
    \left(m^-_{f,\pm}\right)_{ij}\in \left\{-\left(\frac{1}{2}+ \sqrt{\ell^2 + |\mathfrak{Z}_{ij}|^2}\right),-\frac{1}{2}+\sqrt{\ell^2 + |\mathfrak{Z}_{ij}|^2} \right\}.
\end{equation}
These fields correspond to defect operators with dimensions
\begin{equation}    \hat{\Delta}^{\boldsymbol{f}}_{\pm}= 1+ |m_{f, \pm}^{+}|.
\end{equation}
The boundary conditions for fermions are more subtle than those of the vector or scalar fields since the Dirac equation is first order; for a detailed discussion see \cite{Aharony:2010ay,Iqbal:2009fd, Amsel:2008iz}. A Dirac spinor in $AdS_3$ restricts to a pair of chiral fermions on the boundary, with the chirality matrix being the gamma matrix for the radial direction. These two chiral fermions are canonically conjugate to each other, so one of the modes must have Dirichlet boundary conditions, depending on the sign of the mass. The normalizable mode is the one with a positive mass. To be more concrete let us take one of the fermion modes and call it $\xi$. Near $r=0$ the solutions to the Dirac equation have the following asymptotic form
\begin{equation}
    \xi_\pm\sim A_
\pm(x) r^{1\mp m} + B_\pm r^{2\pm m}+\dots,
\end{equation}
so the boundary conditions on the fields depend on the sign of $m$. If $m>0$, the first term is dominant for $\xi_+$ and we should fix $A_+$ at the boundary. The Dirac equation implies that $A_\pm$ is related to $B_\mp$ so that fixing $A_+$ also fixes $B_-$. If $m<0$ we should fix $A_-$. 

The fermion bulk-to-bulk propagator can be expressed in terms of scalar propagators~\cite{Kawano:1999au}
\begin{equation}
    S(x,x')= \sqrt{\frac{r'}{r}}\left[e^{\mu}_{\hat{\alpha}}\,\hat{\gamma}^{\hat{\alpha}} \nabla_\mu+\frac{1}{2}\hat{\gamma}^r+m\right]\left[G_{\Hat{\Delta}-\frac{1}{2}}(x,x')\mathcal{P}_-+ G_{\Hat{\Delta}+\frac{1}{2}}(x,x')\mathcal{P}_+\right],
\end{equation}
where $\mathcal{P}_\pm = \frac{1}{2}(1\pm \hat{\gamma}^r)$.
\section{Perturbative Two-point Function}\label{sec: two pts}
Here, we perturbatively compute the two-point function of single trace operators to compare it with the conformal block expansion in the generic surface defect background. This will allow us to compute bulk-to-defect couplings of BPS operators at leading order in the coupling.
As the simplest case, let us consider the two-point function of the half-BPS scalar single trace operators: 
\begin{align}
    \expval{\mc O_{\Delta_1}(x_1) \mc O_{\Delta_2}(x_2)}
    &= 
    \expval{Y_{I_1}^1\cdots Y_{I_{\Delta_1}}^1\Tr[\phi^{I_1}\cdots \phi^{I_{\Delta_1}}] Y_{J_1}^2\cdots Y_{J_{\Delta_2}}^2\Tr[\phi^{J_1}\cdots \phi^{J_{\Delta_2}}] }
\end{align}
We are interested in the leading-order connected contribution, since self-contractions only contribute to the one-point functions. 
A similar contribution is considered in for example~\cite{deLeeuw:2017dkd, Baerman:2024tql} for the domain wall in $\mc N = 4$ SYM. 
The Feynman diagram for the leading connected contribution is obtained by contracting one of the scalars from each operator as in Fig.~\ref{fig:2pt-leading} with the uncontracted ones giving the contributions of its one-point function in the defect background. 
\begin{figure}[t]
    \centering
    \begin{tikzpicture}
        \foreach \angle in {30,60,...,330} { \draw[] (0,0) -- ({0.5*cos(\angle)}, {0.5*sin(\angle)}) node[draw,shape=circle,fill=black,minimum size=3pt,inner sep=0pt] {}; };
         \foreach \angle in {30,60,...,330} { \draw[] (2,0) -- ({2-0.5*cos(\angle)}, {0.5*sin(\angle)}) node[draw,shape=circle,fill=black,minimum size=3pt,inner sep=0pt] {}; }
         \draw[] (0,0) -- (2,0); 
    \end{tikzpicture}
    \caption{An example of the leading-order diagrams for the two-point functions of the single trace operators. Two vertices at the centers represent each of the single trace operators, connected by a single edge corresponding to the scalar propagator. The black dots represent the one-point functions $\expval{\phi}$ for each Higgs scalar in the surface defect background. }
    \label{fig:2pt-leading}
\end{figure}
This leads to the leading-order contribution of 
\begin{equation}
\begin{aligned}
    &\expval{\mc O_{\Delta_1}(X_1) \mc O_{\Delta_2}(X_2)} \\ 
    &= 
    \frac{1}{\Delta_1\Delta_2}\sum_{m,n} Y_{I_1}^1\cdots Y_{I_{\Delta_1}}^1 Y_{J_1}^2\cdots Y_{J_{\Delta_2}}^2 \\ 
    & \times \expval{\phi^{I_1}}_{i_{\Delta_1} i_1} \expval{\phi^{I_2}}_{i_i i_2} \cdots \expval{\phi^{I_{m-1}}}_{i_{m-2} i_{m-1}}\expval{\phi^{I_{m+1}}}_{i_{m} i_{m+1}}\cdots \expval{\phi^{I_{\Delta_1}}}_{i_{\Delta_1 - 1} i_{\Delta_1}}\\
    & \times \expval{\phi^{J_1}}_{j_{\Delta_2} j_1} \expval{\phi^{J_2}}_{j_i j_2} \cdots \expval{\phi^{J_{n-1}}}_{j_{n-2} j_{n-1}}\expval{\phi^{J_{n+1}}}_{j_{n} j_{n+1}}\cdots \expval{\phi^{J_{\Delta_2}}}_{j_{\Delta_2 - 1} j_{\Delta_2}}\\ 
    & \times \expval{\phi^{I_m}_{i_{m-1} i_{m}}(x_1)\phi^{J_n}_{j_{n-1} j_{n}}(x_2)}
\end{aligned}
\end{equation}
where $\frac{1}{\Delta_1\Delta_2}$ is the symmetry factor, and the $i$ and $j$ indices are in the fundamental of the gauge group. 
For the nonsingular case, the VEVs $\expval{\phi^I}$ take nonzero values in diagonal matrices as \eqref{eq:generic-vev} when $I = 1,2$. 
Let us write the $i$-th diagonal element as $M^I_i$. 
The propagator $\expval{\phi^I \phi^J}$ is the Higgs scalar propagator in the defect background derived in section~\ref{sec:propagators}. 

The important note for this leading-order contribution is that only the diagonal part $i_{m-1} = i_{m}$ and $j_{n-1} = j_{n}$ contributes in the propagator, due to the fact that the one-point functions are diagonal, making $i_1 = i_2 = \cdots = i_{\Delta_1}$ and $j_1 = j_2 = \cdots = j_{\Delta_1}$ for the nonzero terms. 
Furthermore, the propagator $\expval{\phi^I_{ij}\phi^J_{kl}}$ contains the factor $\delta_{il}\delta_{jk}$, which restricts the nonzero terms to have all $i$'s to be the same as $j$'s. 
Thus, the two-point function can be written as 
\begin{align}
    \expval{\mc O_{\Delta_1}(X_1) \mc O_{\Delta_2}(X_2)}_1
    &= 
    \sum_i (Y^1 \circ M_i^1)^{\Delta_1-1}(Y^2 \circ M_i^2)^{\Delta_2-1} \sum_{m,n} Y^1_{I_m} Y^2_{J_n} \expval{\phi^{I_m}_{ii}(x_1)\phi^{J_n}_{ii}(x_2) }
\end{align}
$\circ$ denotes the dot product in the $I = 1,2$ directions only. 
The scalar propagators of the diagonal elements of $\phi$'s are quite simple since the contributions to the effective masses from $\mathfrak Z_{ii}$ all vanish. 
The only remaining mass contributions are due to the KK momenta in the $\psi$ direction. 
Especially, the diagonal parts of the scalars $\phi^1_{ii}$ and $\phi^2_{jj}$ fully decouple
with each other and also from the gauge field fluctuations, and are no longer different from the other scalars. 
The propagator can be thus written as 
\begin{align}
    \expval{\phi^{I_m}_{ii}(x_1)\phi^{J_n}_{ii}(x_2)}
    = \frac{g_{\text{YM}}^2}{4\pi} \frac{1}{r_1 r_2} \delta^{I_m J_n} \sum_{\ell} e^{i\ell \psi} G_{\hat \Delta} (x_1, x_2)
\end{align}
where $G_{\hat \Delta} (x_1, x_2)$ is the $AdS_3$ propagator~\eqref{eq:ads3-propagator} with $\hat \Delta = 1 + |\ell|$. 
The sign of the conformal weight $\hat \Delta$ is chosen so that it is positive to satisfy the unitarity for all $\ell$. 
The geodesic distance $\Tilde \chi$ is related to the cross-ratio $\chi$ as $2/\chi = \Tilde \chi$. 
This matches exactly with the defect conformal symmetry part of the superconformal block by identifying $\hat \Delta$ as the scaling dimension of the internal defect operators in the bulk-defect OPE. 

The product between the polarization vector together with the factor $
\delta^{I_m J_n}$ from the propagator gives the inner product between the polarizations of the two operators
\begin{align}
    Y^1_{I_m} Y^2_{J_n} \delta^{I_m J_n}
     &= (Y^1 \circ Y^1)^{1/2}(Y^2 \circ Y^2)^{1/2} \left(-\frac{\chi_R}{2} + \cos \phi_R \right). 
\end{align}

Lastly, the factor $(Y\circ M_i)^{\Delta - 1}$ can be expanded in terms of the cross-ratios and the VEVs of the scalar fields as  
\begin{align}
    (Y\circ M_i)^{\Delta - 1}
    &= 
    \frac{1}{2^{\Delta - 1}} \frac{(Y \circ Y)^{(\Delta - 1)/2} }{(P \circ P)^{(\Delta - 1)/2}}\left((\beta_i + i \gamma_i)e^{i\eta} + (\beta_i - i \gamma_i)e^{-i\eta}\right)^{\Delta - 1}\nn
    &= 
    \frac{1}{2^{\Delta - 1}} \frac{(Y \circ Y)^{(\Delta - 1)/2} }{(P \circ P)^{(\Delta - 1)/2}}\sum_{k = 0}^{\Delta - 1} {\Delta - 1 \choose k}(\beta_i + i \gamma_i)^{\Delta - k -1} (\beta_i - i \gamma_i)^k e^{i(\Delta - 2k -1)\eta}
\end{align}

Overall, the two-point function in the leading order in terms of the cross ratios looks like 
\begin{align}
    \expval{\mc O_{\Delta_1}(X_1) \mc O_{\Delta_2}(X_2)}
    &= 
    \frac{g_{\text{YM}}^2}{4\pi}\frac{1}{2^{\Delta_1 + \Delta_2}}
    \frac{(Y^1 \circ Y^1)^{\Delta_1/2} (Y^2 \circ Y^2)^{\Delta_2/2}}{(P^1 \circ P^1)^{\Delta_1/2}(P^2 \circ P^2)^{\Delta_2/2}} \nn
    & \times \sum_{k_1 = 0}^{\Delta_1 - 1} \sum_{k_2 = 0}^{\Delta_2 - 1} C(k_1, k_2) e^{ik_1'\eta_1}e^{ik_2'\eta_2} \left(-\frac{\chi_R}{2} + \cos \phi_R \right) \nn
    & 
    \times \sum_\ell e^{i\ell \psi}
    \chi^{-(1+|\ell|)} {}_2F_1\left(\frac{|\ell|+1}{2}, \frac{|\ell|}{2}+1; |\ell| + 1; \frac{4}{\chi^2}\right)
\end{align}
where $C(k_1, k_2)$ is a coefficient 
\begin{align}
    C(k_1, k_2) &=  {\Delta_1 - 1 \choose k_1}{\Delta_2 - 1 \choose k_2} \sum_i(\beta_i + i \gamma_i)^{\Delta_1 + \Delta_2 - k_1 -k_2 -2} (\beta_i - i \gamma_i)^{k_1 + k_2}
\end{align}
and $k'_i = \Delta_i - 2k_i - 1$. 
You can further rewrite it by using $\cos \phi_R e^{i\ell \psi} = \frac{1}{2} \left(e^{i(\ell+1)\psi - i\eta} + e^{i(\ell-1)\psi + i\eta}\right)$ with $\eta = \psi - \phi_R = \eta_1 - \eta_2$ and shifting $\ell$ so that the terms have the same central $SO(2)_t$ charge as  
\begin{align}
    &\expval{\mc O_{\Delta_1}(X_1) \mc O_{\Delta_2}(X_2)}_1\nn
    &= 
    \frac{g_{\text{YM}}^2}{4\pi}\frac{1}{2^{\Delta_1 + \Delta_2 + 1}}
    \frac{(Y^1 \circ Y^1)^{\Delta_1/2} (Y^2 \circ Y^2)^{\Delta_2/2}}{(P^1 \circ P^1)^{\Delta_1/2}(P^2 \circ P^2)^{\Delta_2/2}} \nn
    & \times \sum_{k_1 = 0}^{\Delta_1 - 1} \sum_{k_2 = 0}^{\Delta_2 - 1} \sum_\ell C(k_1, k_2) e^{ik'_1\eta_1 +ik'_2\eta_2} e^{i\ell \psi }\nn 
     &\times \bigg( - \hat{\mathfrak f}_{[\frac{1}{2}, \frac{1}{2}]}^R(\chi_R)\hat{\mathfrak f}_{|\ell|+1}(\chi)
    + e^{-i\eta}\hat{\mathfrak f}_{[0,0]}^R(\chi_R)\hat{\mathfrak f}_{|\ell-1|+1}(\chi)
     + e^{i\eta}\hat{\mathfrak f}_{[0,0]}^R(\chi_R)\hat{\mathfrak f}_{|\ell+1|+1}(\chi)
     \bigg)\\  
        &= \frac{(Y^1 \circ Y^1)^{\Delta_1/2} (Y^2 \circ Y^2)^{\Delta_2/2}}{(P^1 \circ P^1)^{\Delta_1/2}(P^2 \circ P^2)^{\Delta_2/2}} 
     \times \sum_{\hat{\mathcal O}} \sum_{\ell}\mathcal{B}_{\mathcal O_1, \hat{\mathcal O}_{\ell}}(\eta_1)  \mathcal{B}_{\mathcal O_2, \hat{\mathcal O}_{\ell}}(\eta_2)
   \hat{ \mathfrak{F}}_{\hat{\mathcal{S}_\ell}}
\end{align} 

Now, it is obvious how the terms can be packaged as superconformal blocks. 
For a finite $\ell$, the terms coincide with those appearing in the superconformal blocks corresponding to the representation~\eqref{eq:SCB-Sl}.
The $\ell = 0$ terms match with two copies of the superconformal block in the representation~\eqref{eq:SCB-S0}. 
One can see that the relations between the prefactors satisfy the SCWI constraints. The states that appear in the OPE expansion have degeneracy according to the breaking of the gauge symmetry. These states describe point-like open string starting and ending on each of the probe D3-branes.

\section{Discussion}
In this paper we studied correlation functions in the presence of supersymmetric surface operators in $\mathcal{N}=4$ SYM. We used superconformal symmetry to constraint one- and two-point correlation functions of chiral primaries, elaborating on the superconformal block expansion of correlators and the superconformal Ward identities~\cite{Liendo:2016ymz}. Unlike supersymmetric defects of other codimensions, the residual superconformal algebra for a generic surface defect is not enough to fully specify one- and two-point functions and these necessarily become functions of the transverse angular direction and $R$-symmetry polarizations. 
As a check, we quantized the theory around a generic defect by mapping to a supersymmetric vacuum of $\mathcal{N}=4$ SYM on $AdS_3\times S^1$. 
This language should be useful for perturbative calculations since the symmetries of the system are manifest and many techniques for AdS/CFT are applicable. 
We perturbatively computed the leading-order two-point function of chiral primaries at weak coupling and found that it matches our expansion in terms of superconformal blocks.

We also investigated the integrability of half-BPS surface defect operators finding evidence against it for generic defects. This is somewhat expected, since all integrable boundaries known correspond to branes wrapping maximal cycles at strong coupling. The integrable surface defect found in \cite{Dekel:2011ja} is identified with a class of rigid defects sitting at the origin of the moduli space. This set-up is particularly interesting since its quantization necessitates the introduction of additional modes in order to restore the scaling symmetry of the theory. The integrability of the rigid defect system should open the possibility of computing one-point functions of single trace operators at finite coupling. 

We conclude with some future directions:
\begin{itemize}
    \item Two-point functions of chiral primaries and Lorentzian inversion formula: one interesting direction independent of integrability is to apply analytic bootstrap techniques to the GW defect. For instance, bulk three-point functions of chiral primaries are protected as well as their defect one-point functions \cite{Choi:2024ktc}. Can one use Lorentzian inversion \cite{Liendo:2019jpu} to determine bulk-to-defect couplings using this data, and if so, what are the exchanged states in the defect channel? What is the dominant exchange at large transverse spin, or at large charge?
    \item Analytic bootstrap and cohomological sections: Our results are supplementary to the analysis of \cite{Liendo:2016ymz}. Although we did not give a general solution to the superconformal Ward identities for generic multiplets, we believe that there are no clear obstructions. This would be essential for adapting bootstrap techniques to the Gukov-Witten dCFT. For instance, it should be expected that the crossing equations become algebraic equations in cohomological subsectors of the theory \cite{Beem:2013sza,Bianchi:2019sxz}. This would allow to exploit results from supersymmetric localization to compute more general protected correlators \cite{Choi:2024ktc}. Another interesting direction would be to study the coupling dependence for bulk-to-defect two-point functions of BPS operators along the lines of \cite{Baggio:2012rr}. The GW defect two-point function shares many parallels with the bulk four-point function of chiral primaries since their symmetry algebras are analytic continuations of one another. 
    This suggests the possibility that the bulk-to-defect two-point functions of BPS operators are analogous to bulk three-point functions of half-BPS operators. 
    \item Localization and big operators: Recently, it was shown that the one-point functions of chiral primaries in the Gukov-Witten dCFT can be computed using supersymmetric localization~\cite{Choi:2024ktc}. This was used to prove that holographic one-point functions of single-trace operators are finite polynomials in the coupling, with all the contributions coming from tree diagrams at weak coupling. An interesting avenue to explore would be to study correlation functions of determinant, or more Schur polynomial, operators \cite{Corley:2001zk}. These correlators describe intersections of giant graviton branes ending on the probe D3-branes. For these correlators non-planar corrections are important and they encode nonperturbative information in $N$.
    \item Quantization of rigid defects and integrability: the quantization around the rigid defect solutions remains an important problem. We expect that this can be carried out by mapping to a $AdS_3\times S^1$ with a dilaton field. In that case we expect that the conformal coupling of the fields to the background metric should be modified accordingly to realize the underlying superconformal symmetry~\cite{Festuccia:2011ws}. 
    It would also be important to perform checks at strong coupling in the rigid case, especially along the lines of \cite{Jiang:2019xdz}.
    The present article does not fully cover the rigid surface defects discovered in~\cite{Gukov:2008sn}. 
    Although the rigid operators we considered here in the $SU(N)$ gauge theory have the monodoromy whose conjugacy class is a limit of that of the generic cases, the conjugacy classes of the rigid operators in the $SO(N)$ and $Sp(N)$ gauge theories are strictly isolated from the generic ones and hence rigid. 
    They also found some examples of operators with the rigid conjugacy class with diagonal classical profiles for the $SO(N)$ and $Sp(N)$ theories. 
    The relation between the integrability condition and the rigidity of the conjugacy class should be investigated. 
    The extension of our arguments about the integrability to the other semisimple Lie groups is straightforward given the planar limit of these models is essentially the same to that of the $SU(N)$ theory but with additional projections \cite{Caputa:2010ep}.
    One step towards this would be to understand the parity projection beyond the $SU(2)$ sector and compare with the selection rules coming from the boundary state.
    It should also be an interesting question whether this integrability can be discussed from the 4d-2d viewpoint.  
    This is because the $SU(N)$ theory can also have a rigid operator with a rigid conjugacy class, which for the $SU(2)$ case for example, can be constructed by taking a double cover of the target space of the 2d hypermultiplet corresponding to the rigid defect obtained by the $\alpha,\beta,\gamma\rightarrow 0$ limit~\cite{Gukov:2008sn}. 
    In general, such a cover can be taken when the partition of the embedding of $\mathfrak{su}(2)$ to $\mathfrak{su}(N)$ has the greatest common divisor greater than one, and hence the fundamental group of the nilpotent conjugacy class is nontrivial. 
    This construction is unclear from the solely bulk perspective with singular boundary conditions, so taking a cover or not seems not to affect its integrability from our approach.
    This has to be validated from the more obvious 4d-2d view point. 
\end{itemize}

\paragraph{Note Added}
The authors became aware of \cite{Chalabi:2025nbg} which became public soon after the preprint of the present article became available on arXiv. 
The integrability of the Gukov-Witten type surface defects is also discussed there.
They made our result more precise, concluding that the rigid surface defect is integrable in all bosonic sectors only if the $\mathfrak{su}(2)$ representations of the field profiles are two-dimensional.

\paragraph{Acknowledgments}
The authors thank Shota Komatsu for many fruitful discussions since the early stage of this work, providing us some computer programs for numerical computations, and giving us helpful comments on the preliminary version of this manuscript. We also thank Changha Choi for discussions. AH would like to thank Marius de Leeuw for discussions and comments on the preliminary version of this manuscript. HK thanks David Berenstein, David Grabovsky, and Yu Nakayama for fruitful discussions. Part of the HK's work was done during the visit at Yukawa Institute for Theoretical Physics funded by the "Atom-type Researcher" program. 
He thanks them for their hospitality and support.
The work of AH is supported by ERC-2022-CoG - FAIM 101088193. 
HK is partly supported by the Department of Energy under Award No. DE-SC0019139. 

\appendix

\section{Harmonic Superspace Formalism \label{sec:superspace}}
The harmonic superspace formalism is utilized for studying various types of the superconformal field theories, for example in~\cite{Dolan:2004mu, Doobary:2015gia, Liendo:2015cgi, Liendo:2016ymz}. 
The setup for the surface defect configuration under our consideration is already provided in~\cite{Liendo:2016ymz}. 
The spacetime coordinates and the $SO(6)_R$ polarization can be packaged as a $(2|2)$ matrix together with the Grassmann odd coordinates as 
\begin{align}
    X^{A\dot A} = \begin{pmatrix}
        x^{\alpha\dot \alpha} & \theta^{\alpha \dot a}\\
        \bar \theta^{a \dot \alpha} & y^{a\dot a}
    \end{pmatrix}. 
\end{align}
The uppercase indices run from $1, ..., 4$, and the lowercase indices run from $1,2$. 
$x^{\alpha\dot \alpha}$ is the ordinary four-dimensional coordinates in the spinor representation.
$y^{a\dot a}$ specifies the polarization. 
The null vector $Y^I$ ($I = 1,2,...,6$) characterizing the 1/2-BPS operators is related to $y^{a\dot a}$ as 
\begin{align}\label{eq:r-symm-nonlinear}
    Y^I = \left (y^\mu, \frac{1}{2}(1 - y^2), \frac{i}{2}(1 + y^2)\right), \quad 
    y^{a \dot a} = y^\mu (\sigma_\mu)^{a\dot a}, 
\end{align}
where $\sigma_\mu$ are the Pauli matrices. 
$\theta$ and $\bar \theta$ are the odd coordinates.
The $PSU(2, 2|4)$ group elements act on the vector space of $X = \begin{pmatrix}
    A & B\\ C & D
\end{pmatrix}$ as 
\begin{align}\label{eq:psu224-superspace}
    g \circ X = (AX + B)(CX + D)^{-1}. 
\end{align}

$X$ can be split into the directions parallel to the surface and perpendicular to the surface as $X = X_S + X_\perp$, where 
\begin{align}
    X_S = \chi_S \otimes \begin{pmatrix}
        1 & 0 \\ 0 & 0 
    \end{pmatrix}
    + \bar \chi_S \otimes \begin{pmatrix}
        0 & 0 \\ 0 & 1
    \end{pmatrix}, 
    \quad 
    X_\perp  = \chi_\perp \otimes \begin{pmatrix}
        0 & 1 \\ 0 & 0 
    \end{pmatrix}
    + \bar \chi_\perp \otimes \begin{pmatrix}
        0 & 0 \\ 1 & 0
    \end{pmatrix}. 
\end{align}
The elements of the residual subgroup $PSU(1,1|2)^2\ltimes SO(2)_t$ acts just as eq.~\eqref{eq:psu224-superspace} with the elements
\begin{align}
    g = \begin{pmatrix}
        A & B \\ C & D
    \end{pmatrix}
    = 
    \begin{pmatrix}
        a & b \\ c & d
    \end{pmatrix}
    \otimes 
    \begin{pmatrix}
        1 & 0 \\ 0 & 0 
    \end{pmatrix}
    + \begin{pmatrix}
        \bar a & \bar b \\ \bar c & \bar d
    \end{pmatrix}
    \otimes \begin{pmatrix}
        0 & 0 \\ 0 & 1
    \end{pmatrix}. 
\end{align}
The lowercase variables ($\chi$'s and the roman alphabets) are all $(1|1)$ matrices, and the superdeterminants of $\begin{pmatrix} a & b \\ c & d \end{pmatrix}$ and $\begin{pmatrix} \bar a & \bar b \\ \bar c & \bar d \end{pmatrix}$ are both unity. 
It is worth noting the specific transformations of the superspace under this subgroup:
\begin{itemize}
    \item Supertranslations ($A = D = 1_{(2|2)}$, $B = \diag(b, \bar b)$, $C = 0$)
    \begin{align}
        (\chi_S, \bar \chi_S, \chi_\perp, \bar \chi_\perp) 
        \mapsto (\chi_S + b, \bar \chi_S + \bar b, \chi_\perp, \bar \chi_\perp) 
    \end{align}
    \item Superrotations ($A = \diag(a, \bar a)$, $D = \diag(d, \bar d)$, $B = C = 0$)
    \begin{align}
        (\chi_S, \bar \chi_S, \chi_\perp, \bar \chi_\perp) 
        \mapsto
        (a\chi_S d^{-1}, \bar a \bar \chi_S \bar d^{-1}, a \chi_\perp \bar d^{-1}, \bar a \bar \chi_\perp d^{-1}) 
    \end{align}
    \item Superspecial conformal transformations 
    \begin{itemize}
        \item with $\bar c = 0$ ($A = D = 1_{(2|2)}$, $C = \diag(c, 0)$, $B = 0$)
    \begin{align}
        &(\chi_S, \bar \chi_S, \chi_\perp, \bar \chi_\perp) \nn
        &\mapsto
        (\chi_S(1 + c \chi_S)^{-1}, \bar \chi_S - \bar \chi_\perp (1 + c \chi_S)^{-1} c \chi_\perp, (1 + \chi_Sc )^{-1}\chi_\perp, \bar \chi_\perp(1 + c \chi_S)^{-1}) 
    \end{align}
    \item with $c = 0$ ($A = D = 1_{(2|2)}$, $C = \diag(0, \bar c)$, $B = 0$)
    \begin{align}
        &(\chi_S, \bar \chi_S, \chi_\perp, \bar \chi_\perp) \nn
        &\mapsto
        (\chi_S - \chi_\perp (1 + \bar c \bar \chi_S)^{-1} \bar c \bar \chi_\perp, \bar \chi_S(1 + \bar c \bar \chi_S)^{-1}, \chi_\perp(1 + \bar c \bar \chi_S)^{-1}, (1 + \bar \chi_S \bar c )^{-1}\bar \chi_\perp) 
    \end{align}
    \end{itemize}
    Notice that on the surface $\chi_\perp = \bar \chi_\perp = 0$, the superspecial conformal transformations do not mix $\chi_S$ and $\bar \chi_S$. 
\end{itemize}

\section{Calculation of conformal blocks \label{sec:conformal-blocks}}
The bulk-defect OPEs decompose the bulk operators to the sum of the defect operators. 
This can be seen from the radial quantization around some point on the defect and the defect vacuum $\ket{\hat \Omega}$ is defined as a state on a sphere around the point~\cite{Billo:2016cpy}.
The action of some bulk operator in this case is deforming the sphere so that the operator goes from its exterior to interior, or vice versa. 
The two-point function can be related to the bulk-defect OPEs by inserting the resolution of the identity as 
\begin{align}
    \expval{\mc O_1 \mc O_2}
    &= \mel{\hat \Omega}{\mc O_1 \mc O_2}{\hat \Omega}
    = \sum_{\hat \alpha} \mel{\hat \Omega}{\mc O_1}{\hat \alpha}\mel{\hat \alpha}{\mc O_2}{\hat \Omega}, 
\end{align}
where $\hat \alpha$ are the defect operators. 
Let $\hat J_i$ be a generator of the defect symmetry, then one finds that each contribution in this conformal block expansion is an eigenfunction of the Casimir operator of the symmetry: 
\begin{align}\label{eq:bulk-defect-casimir-eqns}
    \hat J_{1i} \hat J^i_1 \expval{\mc O_1 \mc O_2}
    \supset \mel{\hat \Omega}{[\hat {\mc J}_i, [\hat {\mc J}^i, \mc O_1]]}{\hat \alpha}\mel{\hat \alpha}{\mc O_2}{\hat \Omega}
    = C_{\hat \alpha} \mel{\hat \Omega}{\mc O_1}{\hat \alpha}\mel{\hat \alpha}{\mc O_2}{\hat \Omega}, 
\end{align}
where $\hat J_{1i}$ is the symmetry generator acting on the first insertion point, $\hat {\mc J}_i$ is the operator corresponding to this generator, and $C_{\hat \alpha}$ is the value of the quadratic Casimir characterizing the representation in which $\ket {\hat \alpha}$ resides, given that $|\hat \Omega\rangle$ is invariant under this symmetry, i.e. $\hat J_i |\hat \Omega\rangle = 0$.

This means that the defect conformal block $\mathfrak{f}_{\hat \Delta}(\chi) \mathfrak{f}^R_{[\frac{n}{2}, \frac{n}{2}]}(\chi_R) \mathfrak{h}_j (\psi)$ must be an eigenfunction of the quadratic Casimir of the residual bosonic symmetry. 
The quadratic Casimirs of each of  $SO(2,2)$, $SO(4)_R$, and $SO(2)_t$ are 
\begin{align}
    &\hat L^2 \equiv \frac{1}{2} J_{1AB}J_1^{AB}, \quad 
    \hat L_R^2 \equiv \frac{1}{2} R_{1AB} R_1^{AB}, \quad 
    \hat S^2 \equiv \left(J_{112} + R_{112}\right)^2
\end{align}
respectively, where $J_{iMN}$ are the generators of the $SO(1, 5)$ Lorentz symmetry in the embedded six-dimensional space 
\begin{align}
    J_{iMN} = P_{iM} \pdv{}{P_i^N} - P_{iN} \pdv{}{P_i^M}
\end{align}
acting on the $i$-th insertion point, and $R_{iMN}$ are the generators of the $SO(6)_R$ R-symmetry 
\begin{align}
    R_{iMN} = Y_{iM} \pdv{}{Y_i^N} - Y_{iN} \pdv{}{Y_i^M}
\end{align}
again acting on the $i$-th point. 
Each differential operator has a nontrivial dependence on only one of the cross-ratios. 
So, the Casimir equations can be written separately for each bosonic subgroup
\begin{align}
    &(\hat L^2 + \hat \Delta(\hat \Delta-2))\mathfrak{f}_{\hat \Delta}(\chi) = 0,  \\
    &(\hat L_R^2 + n(n + 2))\mathfrak{f}^R_{[\frac{n}{2}, \frac{n}{2}]}(\chi_R)  = 0, \\
    &(\hat S^2 + j^2) \mathfrak{h}_j (\psi) = 0. 
\end{align}
Refer to Sec.~\ref{sec:bosonic-algebra} for the discussion on the dependence of the values of the Casimirs on the quantum numbers.  
Note that all the Casimir operators act on only one of the two inserted points, let us say $X_1$, as in \eqref{eq:bulk-defect-casimir-eqns}.

The action of the Casimir of $SO(2,2)$ can be rewritten in the same manner in terms of the cross-ratios as derived in~\cite{Billo:2016cpy}: 
\begin{align}
    \left[(4 - \chi^2)\pdv{}{\chi^2} - 3 \chi \pdv{}{\chi} + \hat \Delta(\hat \Delta-2)\right]\mathfrak{f}_{\hat \Delta}(\chi) = 0.
\end{align}
Solving this equation, the eigenfunction for the conformal symmetry part is found to be 
\begin{align} 
    \mathfrak{f}_{\hat \Delta}(\chi)
    =  
    \chi^{-\hat \Delta} {}_2 F_1 \left(\frac{\hat \Delta + 1}{2}, \frac{\hat \Delta}{2}; \hat \Delta; \frac{4}{\chi^2}\right). 
\end{align}
The conformal block for the $SO(4)_R$ part can be obtained by replacing $\chi$ with $\chi_R$ and $\hat \Delta$ with $-n$. 
In this case, it is a polynomial for nonnegative integer $n$ as expected: 
\begin{align} 
     \hat f^R_{[\frac{n}{2}, \frac{n}{2}]}(\chi_R)
    &= 
    \chi_R^{n} {}_2 F_1 \left(\frac{-n + 1}{2}, -\frac{n}{2}; n; \frac{4}{\chi_R^2}\right)
    = \sum_{m = 0}^{\lfloor n/2 \rfloor} (-1)^m {\lfloor n/2 \rfloor \choose m} 
    \frac{(n/2 (+1/2))_m}{(n)_m}4^{2m} \chi_R^{n - 2m}. 
\end{align}

For the mixed $SO(2)_t$ part, the Casimir equation reduces to 
\begin{align}
    &\left[(1 - \cos^2 \psi)\pdv{^2}{\cos\phi_+^2} - \cos\psi \pdv{}{\cos\psi}  + \hat j^2 \right] ={\mathfrak h}_{j}(\psi) = 0
    \implies  
    \left[\pdv{^2}{\psi^2} + \hat j^2 \right] {\mathfrak h}_{j}(\psi)
    = 0
\end{align}
The solution is\footnote{$e^{-i j\psi}$ is also a valid solution of this equation, but since the representations with $j$ and $-j$ should always appear symmetrically, we choose to absorb this into the block with ${\mathfrak h}_{-j}(\psi)$. }
\begin{align}
    {\mathfrak h}_{j}(\psi)
    =  e^{i j\psi}
\end{align}

We can perform a similar calculation to determine a general form of the $\eta$ dependent part, even though the full decomposition cannot be determined due to the lack of symmetry. 
This $\eta$ dependence is due to the fact that the defect vacuum may have a nonzero charge under the broken part of the $SO(2)\times SO(2)_R$ rotations. 
The Casimir of this broken $SO(2)$ acting on the $i$-th insertion point is 
\begin{align}
    \hat {S'}_i^2 = (J_{i12} - R_{i12})^2
\end{align}
From the two-dimensional theory point of view after dimensional reduction in the defect directions, this corresponds to the boundary condition characterized by $\pi_1((\mathbb R^2\setminus \{0\}) \times (\mathbb R^2\setminus \{0\}))/\pi_1(\mathbb R^2\setminus \{0\}) \simeq \mathbb Z$ around the origin where the defect is inserted. 
Let us write the defect state in the general form $|\hat \Omega\rangle = \sum_\ell c_\ell |\hat \Omega_\ell\rangle$. 
Action of the Casimir operator on the two-point function evaluated with this vacuum state gives
\begin{align}
   \hat {S'}_i^2 \expval{\mc O_1 \mc O_2}
    &= 
    \sum_{\ell, \ell'} \sum_{\hat \alpha}c_\ell c_{\ell'} \mel{\hat \Omega_\ell}{[\hat S', [\hat S', \mc O_1]]}{\hat \alpha}\mel{\hat \alpha}{\mc O_2}{\hat \Omega_{\ell'}}\nn 
    &= 
    \sum_{\ell, \ell'} \sum_{\hat \alpha}c_\ell c_{\ell'} (\ell - \hat j')^2 \mel{\hat \Omega_\ell}{\mc O_1}{\hat \alpha}\mel{\hat \alpha}{\mc O_2}{\hat \Omega_{\ell'}}
\end{align}
given that $\hat S' |\hat \Omega\rangle = \ell |\hat \Omega\rangle $. 
$\hat j'$ is the charge of $\ket{\hat \alpha}$ under this broken $SO(2)$. 
The action of $\hat {S'}_2^2$ is in the similar manner. 
$\hat {S'}_i^2$ can be translated to the second derivative $4\del^2/\del \eta_i^2$.
This leads to that each contribution in the decomposition is $\propto C_{\ell, +} e^{i(\ell - \hat j') \eta_i/2} + C_{\ell, -} e^{-i(\ell - \hat j') \eta_i/2}$.
Overall, one can obtain the general form for the part depending on $\eta_1, \eta_2$ for a specific exchanged operator charged with $\hat j'$ under the broken $SO(2)$
\begin{align}
    \mathcal{B}_{1, \hat{\rho}}(\eta_1)\mathcal{B}_{2, \hat{\rho}}(\eta_2)C^{\hat{\rho}}_{\hat{\Delta}, n}\,(\eta)
    & = \sum_{\ell, \ell'} (c^{\hat{\rho}}_{\hat{\Delta}, n, \ell}e^{i(\ell - \hat j') \eta_1/2} + \Tilde c^{\hat{\rho}}_{\hat{\Delta}, n, \ell}e^{-i(\ell - \hat j') \eta_1/2})\nn
    & \times (c^{\hat{\rho}}_{\hat{\Delta}, n, \ell'} e^{i(\ell' - \hat j') \eta_2/2} + \Tilde c^{\hat{\rho}}_{\hat{\Delta}, n, \ell'}  e^{-i(\ell' - \hat j') \eta_2/2})
\end{align}

\section{Representations of $PSU(1,1|2)^2\ltimes SO(2)_t$ \label{sec:residual-reps}}
Here, we study the classification of the representations of the residual symmetry $PSU(1,1|2)^2\ltimes SO(2)_t$. 
We start by reviewing the classification of long and short multiplets of the complexified $\mathfrak{sl}(2|2)$ following~\cite{Zhang:2004qx}. 
The unitary representations of $\mathfrak{sl}(2|2)$ have finite dimensions. 
Its extension to the unitary representations of $\mathfrak{sl}(1,1|2)$, which are infinite dimensional, appearing in our discussion is fairly straightforward. 
A defect symmetry representation is constructed by combining a pair of $\mathfrak{psl}(1,1|2)$ multiplets with the same central charge under $\mathfrak{so}(2)_t$. 
We study how it decomposes into representations of the bosonic subalgebra $\mathfrak{so}(2,2)\oplus \mathfrak{so}(4)_R \oplus \mathfrak{so}(2)_t$ of the defect symmetry to find the combinations of the quantum numbers appearing in the superconformal block decomposition.

\subsection{Finite-dimensional long representations of $\mathfrak{sl}(2|2)$}
Let us first review the finite-dimensional long (non-BPS) representations of $\mathfrak{sl}(2|2)$. 
In this case, the behavior is the same for the massless and massive cases. 
The raising fermionic generators ${\mathfrak{Q}^\alpha}_a$ are in the representation $[\frac{1}{2}, \frac{1}{2}]$\footnote{We denote the Dynkin labels of $\mathfrak g_0 = \mathfrak{sl}(2)\oplus \mathfrak{sl}(2)$ with square brackets $[\cdot, \cdot]$, the long representations of the whole superalgebra $\mathfrak{(p)sl}(2|2)$ with curly brackets $\{\cdot, \cdot\}$, and the short representations with angular brackets $\langle \cdot, \cdot \rangle$. 
We take the convention in which the labels $j_1, j_2$ of $[j_1, j_2]$ can take values as either integer or half-integer values, or spins. 
They are half of the Dynkin labels of $\mathfrak g_0$. 
} of the bosonic subalgebra $\mathfrak g_0 $, so are the lowering generators ${\mathfrak{S}^{\dot a}}_{\dot \alpha}$. 
We label the long supermultiplet as $\{j_1, j_2\}$ whose top component has the $\mathfrak g_0= \mathfrak{sl}(2)\oplus \mathfrak{sl}(2)$ representation $[j_1, j_2]$. 
The commutation relation 
\begin{align}
    [{(\mathfrak{L}_i)^{\alpha}}_{\dot \alpha}, {(\mathfrak{Q}_i)^{\beta}}_{b}\}]
    = 2\delta^\beta_{\dot \alpha} {(\mathfrak{Q}_i)^{\alpha}}_{b} - \delta^\alpha_{\dot \alpha} {(\mathfrak{Q}_i)^{\beta}}_{b} , \quad 
    [{(\mathfrak{R}_i)^{\dot a}}_a, {(\mathfrak{Q}_i)^{\beta}}_{b}\}]
    = -2\delta_{b \dot a} {(\mathfrak{Q}_i)^{\beta}}_{a} + \delta^a_{\dot a} {(\mathfrak{Q}_i)^{\beta}}_{b}
\end{align}
means that the ${(\mathfrak{Q}_i)^{\alpha}}_{a}$ operators maps the state $\ket{s_1, s_2}$ labeled with the eigenvalues of the Cartan generators in $[j_1, j_2]$ to a linear combination of the states in other representations labeled with $\ket{s_1 + \frac{(-1)^{\alpha -1}}{2}, \, s_2 + \frac{(-1)^a}{2}}$. 
This suggests that four independent odd operators $z_i$, $i = 1,2,3,4$ that map a $\mathfrak g_0$ multiplet $[j_1, j_2]$ to another multiplet $[j_1 \pm \frac{1}{2}, j_2\pm \frac{1}{2}]$, respectively, can be constructed linearly from ${(\mathfrak{Q}_i)^{\alpha}}_{a}$ together with bosonic operators.
The nilpotency of the supercharges tells us that all $\mathfrak g_0$ multiplets are exhausted by acting $z_4^{n_4}z_3^{n_3}z_2^{n_2}z_1^{n_1}$ ($n_i = 0,1$) on the top component of the supermultiplet $\{m,n\}$.
The long representation is thus composed of the finite web of the $\mathfrak g_0$ representations as Fig.~\ref{fig:psu22-long}. 
\begin{figure}[t]
    \centering
    \begin{tikzpicture}
    \useasboundingbox(-5, -2.2) rectangle (5,2.2); 
        \pgfmathsetmacro{\d}{0.7}
        \node[draw=gray, thick, inner sep=2pt] (mn) {$[j_1,j_2]$}; 
        \node[above right = \d of mn] (m1n1) {\scriptsize $[j_1+\frac{1}{2},j_2+\frac{1}{2}]$};
        \node[above left = \d of mn] (m-1n1) {\scriptsize $[j_1-\frac{1}{2},j_2+\frac{1}{2}]$};
        \node[below right = \d of mn] (m1n-1) {\scriptsize $[j_1+\frac{1}{2},j_2-\frac{1}{2}]$};
        \node[below left = \d of mn] (m-1n-1) {\scriptsize $[j_1-\frac{1}{2},j_2-\frac{1}{2}]$};
        \node[above = {2*\d} of mn] (mn2) {\scriptsize $[j_1,j_2+1]$};
        \node[right = {4*\d} of mn] (m2n) {\scriptsize $[j_1+1,j_2]$};
        \node[left = {4*\d} of mn] (m-2n) {\scriptsize $[j_1-1,j_2]$};
        \node[below = {2*\d} of mn] (mn-2) {\scriptsize $[j_1,j_2-1]$};
        \draw[->] (mn) -- node[midway, below left = -0.1] {$z_1$} (m1n-1) ; 
        \draw[->] (mn) -- node[midway, above left = -0.1] {$z_2$} (m1n1) ; 
        \draw[->] (mn) -- node[midway, above right = -0.1] {$z_3$} (m-1n1) ; 
        \draw[->] (mn) -- node[midway, below right = -0.1] {$z_4$} (m-1n-1) ; 
        \draw[->] (m1n-1) -- node[midway, above] {\tiny 12} (m2n) ; 
        \draw[->] (m1n-1.120) -- node[midway, above] {\tiny 13} (mn.350) ; 
        \draw[->] (m1n-1) -- node[midway, above] {\tiny 14} (mn-2) ; 
        \draw[->] (m2n) -- node[midway, above] {\tiny 123} (m1n1) ; 
        \draw[->] (m2n.250) -- node[midway, below] {\tiny 124} (m1n-1.20) ; 
        \draw[->] (m1n1.250) -- node[midway, right = 0.1] {\tiny 24, 1234} (mn.10) ; 
        \draw[->] (m1n1) -- node[midway, above] {\tiny 23} (mn2) ; 
        \draw[->] (mn2) -- node[midway, above] {\tiny 234} (m-1n1) ;
        \draw[->] (m-1n1) -- node[midway, above] {\tiny 34} (m-2n) ;
        \draw[->] (mn.190) -- node[midway, above] {\tiny 134} (m-1n-1.60) ;
    \end{tikzpicture}
    \caption{The decomposition structure of the long representation $\{m,n\}$ of $\mathfrak{sl}(2|2)$ to the representations of the bosonic subalgebra $\mathfrak g_0 = \mathfrak{sl}(2)\oplus \mathfrak{sl}(2)$. The boxed component is the top component. The sequence of the numbers labeling the arrows denote which odd operators are turned on, e.g. $234$ means the action with $z_4 z_3 z_2$. }
    \label{fig:psu22-long}
\end{figure}

\subsection{Finite-dimensional short representations of $\mathfrak{sl}(2|2)$}
The shortening conditions of the supermultiplet with the top component $[j_1,j_2]$ are~\cite{Zhang:2004qx}
\begin{align}\label{eq:psu22-shortening}
    C = \begin{cases}
        \pm 2\left(j_1 - j_2\right) & \textbf{(Type I)}  \\
        \pm 2\left(j_1 + j_2 + 1\right) & \textbf{(Type II)}
    \end{cases}. 
\end{align}

If the first Type~I condition $C = 2\left(j_1 - j_2\right) $ condition is satisfied with massive $C\neq 0$, then the action of $z_2$ annihilates (see Fig.~\ref{fig:psu22-short-z2}).
We call this short multiplet $\expval{j_1, j_2}_{\text{I}+}$.
\begin{figure}[t]
    \centering
    \begin{subfigure}[b]{0.48\textwidth}
    \centering 
        \begin{tikzpicture}[baseline]
        \useasboundingbox(-5, -2.2) rectangle (3,2.2);
        \pgfmathsetmacro{\d}{0.7}
        \node[anchor=base, draw=gray, thick, inner sep=2pt] (mn) {$[j_1, j_2]$}; 
        \node[above left = \d of mn] (m-1n1) {\scriptsize $[j_1-\frac{1}{2},j_2+\frac{1}{2}]$};
        \node[below right = \d of mn] (m1n-1) {\scriptsize $[j_1+\frac{1}{2},j_2-\frac{1}{2}]$};
        \node[below left = \d of mn] (m-1n-1) {\scriptsize $[j_1-\frac{1}{2},j_2-\frac{1}{2}]$};
        \node[left = {4*\d} of mn] (m-2n) {\scriptsize $[j_1-1,j_2]$};
        \node[below = {2*\d} of mn] (mn-2) {\scriptsize $[j_1,j_2-1]$};
        \draw[->] (mn) -- node[midway, below left=] {\tiny 1} (m1n-1) ; 
        \draw[->] (mn) -- node[midway, above right] {\tiny 3} (m-1n1) ; 
        \draw[->] (mn) -- node[midway, below right] {\tiny 4} (m-1n-1) ;  
        \draw[->] (m1n-1.120) -- node[midway, above] {\tiny 13} (mn.350) ; 
        \draw[->] (m1n-1) -- node[midway, above] {\tiny 14} (mn-2) ; 
        \draw[->] (m-1n1) -- node[midway, above] {\tiny 34} (m-2n) ;
        \draw[->] (mn.190) -- node[midway, above] {\tiny 134} (m-1n-1.60) ;
    \end{tikzpicture}
    \caption{\label{fig:psu22-short-z2}}
    \end{subfigure}
    \hfill 
    \begin{subfigure}[b]{0.48\textwidth}
    \centering 
        \begin{tikzpicture}[baseline]
        \useasboundingbox(-3, -2.2) rectangle (5,2.2);
        \pgfmathsetmacro{\d}{0.7}
        \node[anchor=base,draw=gray, thick, inner sep=2pt] (mn) {$[j_1,j_2]$}; 
        \node[above right = \d of mn] (m1n1) {\scriptsize $[j_1+\frac{1}{2},j_2+\frac{1}{2}]$};
        \node[above left = \d of mn] (m-1n1) {\scriptsize $[j_1-\frac{1}{2},j_2+\frac{1}{2}]$};
        \node[below right = \d of mn] (m1n-1) {\scriptsize $[j_1+\frac{1}{2},j_2-\frac{1}{2}]$};
        \node[above = {2*\d} of mn] (mn2) {\scriptsize $[j_1,j_2+1]$};
        \node[right = {4*\d} of mn] (m2n) {\scriptsize $[j_1+1,j_2]$};
        \draw[->] (mn) -- node[midway, below left = -0.1] {\tiny 1} (m1n-1) ; 
        \draw[->] (mn) -- node[midway, above left = -0.1] {\tiny 2} (m1n1) ; 
        \draw[->] (mn) -- node[midway, above right = -0.1] {\tiny 3} (m-1n1) ; 
        \draw[->] (m1n-1) -- node[midway, above] {\tiny 12} (m2n) ; 
        \draw[->] (m1n-1.120) -- node[midway, above] {\tiny 13} (mn.350) ; 
        \draw[->] (m2n) -- node[midway, above] {\tiny 123} (m1n1) ; 
        \draw[->] (m1n1) -- node[midway, above] {\tiny 23} (mn2) ; 
    \end{tikzpicture}
    \caption{\label{fig:psu22-short-z4}}
    \end{subfigure}
    \caption{The decomposition structure of the massive short representations $\expval{j_1 , j_2}_{\text{I}\pm}$ with (a) $C = 2\left(j_1 - j_2\right)$ and (b) $C = -2\left(j_1 - j_2\right)$.}
    \label{fig:psu22-short-type1}
\end{figure}
There are some special cases with small values of $j_1$ or $j_2$. 
For example, in $\expval{0,j_2}_{\text{I}+}$, only the nonnegative Dynkin label representations remain, plus the action of $z_3 z_1$ annihilates the top component. 
Hence, the branching rule consists of three bosonic components as in Fig.~\ref{fig:psu22-short-z2-m0}. 
Similarly, $\expval{j_1,0}_{\text{I}+}$ is as in Fig.~\ref{fig:psu22-short-z2-n0}. 
The representations with $j_1 = \frac{1}{2}$ or $j_2 = \frac{1}{2}$ can be obtained by simply removing the $[j_1 - 1,j_2]$ or $[j_1, j_2-1]$ multiplet, respectively. 
\begin{figure}
    \centering
    \begin{subfigure}[b]{0.45\linewidth}
    \centering 
\begin{tikzpicture}[baseline=(mn.base)]
    \useasboundingbox(-2, -2.2) rectangle (2,2.2);
        \pgfmathsetmacro{\d}{0.7}
        \node[anchor=base,draw=gray, thick, inner sep=2pt] (mn) {$[0,j_2]$}; 
        \node[below right = \d of mn] (m1n-1) {\scriptsize $[\tfrac{1}{2},j_2-\frac{1}{2}]$};
        \node[below = {2*\d} of mn] (mn-2) {\scriptsize $[0,j_2-1]$};
        \draw[->] (mn) -- node[midway, below left=] {\tiny 1} (m1n-1) ; 
        \draw[->] (m1n-1) -- node[midway, above] {\tiny 14} (mn-2) ; 
    \end{tikzpicture}        \caption{\label{fig:psu22-short-z2-m0}}
    \end{subfigure}
     \hfill 
    \begin{subfigure}[b]{0.45\linewidth}
    \centering 
    \begin{tikzpicture}[baseline=(mn.base)]
    \useasboundingbox(-3, -2.2) rectangle (1,2.2);
        \pgfmathsetmacro{\d}{0.7}
        \node[anchor=base,draw=gray, thick, inner sep=2pt] (mn) {$[j_1,0]$}; 
        \node[above left = \d of mn] (m-1n1) {\scriptsize $[j_1-\frac{1}{2},\tfrac{1}{2}]$};
        \node[left = {4*\d} of mn] (m-2n) {\scriptsize $[j_1-1, 0]$};
        \draw[->] (mn) -- node[midway, above right] {\tiny 3} (m-1n1) ; 
        \draw[->] (m-1n1) -- node[midway, above] {\tiny 34} (m-2n) ;
    \end{tikzpicture}
    \caption{\label{fig:psu22-short-z2-n0}}
    \end{subfigure}
    \caption{Special cases of (a) $\expval{j_1 , 0}_{\text{I}+}$ or (b) $\expval{0 , j_2}_{\text{I}+}$ for the representations satisfying the shortening condition $C = 2 \left(j_1 - j_2\right)$. }
    \label{fig:psu22-short-type11-mn0}
\end{figure}

The massive Type~I representation with $C = -2\left(j_1 - j_2\right)$ has the multiplets annihilated by $z_4$ (Fig.~\ref{fig:psu22-short-z4}). 
We call it $\expval{j_1 , j_2}_{\text{I}-}$.
In contrary to the $\expval{0,j_2}_{\text{I}+}$ and $\expval{j_1,0}_{\text{I}+}$ multiplets, nothing special happens for the cases of $j_1 = 0$ or $j_2 = 0$ rather than missing the multiplets with possibly negative labels. 

For a massless $C=0$ representation, both of the Type~I conditions are satisfied if $j = j_1 = j_2$.
We call this $\expval{j, j}_{\text{I}\pm}$. 
It suggests that both $z_2$ and $z_4$ annihilate the components, and it is indeed the case (Fig.~\ref{fig:psu22-short-z24}). 
\begin{figure}[t]
    \centering
        \begin{tikzpicture}[baseline]
        \pgfmathsetmacro{\d}{0.7}
        \node[anchor=base,draw=gray, thick, inner sep=2pt] (mn) {$[j,j]$}; 
        \node[above left = \d of mn] (m-1n1) {\scriptsize $[j-\frac{1}{2}, j+\frac{1}{2}]$};
        \node[below right = \d of mn] (m1n-1) {\scriptsize $[j+\frac{1}{2},j-\frac{1}{2}]$};
        \draw[->] (mn) -- node[midway, below left=] {\tiny 1} (m1n-1) ; 
        \draw[->] (mn) -- node[midway, above right] {\tiny 3} (m-1n1) ; 
        \draw[->] (m1n-1.120) -- node[midway, above] {\tiny 13} (mn.350) ; 
    \end{tikzpicture}
    \caption{The multiplet structure of a massless representation $\expval{j, j}_{\text{I}\pm}$ with $j = j_1 = j_2$ satisfying both of the Type~I conditions.}
    \label{fig:psu22-short-z24}
\end{figure}
This class of short representations includes the one-dimensional trivial representation $\expval{0,0}_{\text{I}\pm}$. 

The right-hand side of the Type~II shortening conditions $C = \pm 2\left(j_1 + j_2 + 1\right)$ is strictly nonzero for nonnegative $j_1$ and $j_2$, so it may be applicable only for massive $C \neq 0$ representations for finite representations. 
If the first condition $C = 2\left(j_1 + j_2 + 1\right)$ is satisfied, the top component is annihilated by $z_1$ (Fig.~\ref{fig:psu22-short-z1}). 
We call this multiplet $\expval{j_1, j_2}_{\text{II}+}$. 
\begin{figure}[t]
    \centering
    \begin{subfigure}{0.48\textwidth}
    \centering
        \begin{tikzpicture}[baseline=(mn.base), yshift=2cm]
        \useasboundingbox(-5, -2.2) rectangle (3,2.2);
        \pgfmathsetmacro{\d}{0.7}
        \node[draw=gray, thick, inner sep=2pt] (mn) {$[j_1, j_2]$}; 
        \node[above right = \d of mn] (m1n1) {\scriptsize $[j_1+\frac{1}{2},j_2+\frac{1}{2}]$};
        \node[above left = \d of mn] (m-1n1) {\scriptsize $[j_1-\frac{1}{2},j_2+\frac{1}{2}]$};
        \node[below left = \d of mn] (m-1n-1) {\scriptsize $[j_1-\frac{1}{2},j_2-\frac{1}{2}]$};
        \node[above = {2*\d} of mn] (mn2) {\scriptsize $[j_1,j_2+1]$};
        \node[left = {4*\d} of mn] (m-2n) {\scriptsize $[j_1-1,j_2]$};
        \draw[->] (mn) -- node[midway, above left = -0.1] {\tiny  2} (m1n1) ; 
        \draw[->] (mn) -- node[midway, above right = -0.1] {\tiny 3} (m-1n1) ; 
        \draw[->] (mn) -- node[midway, below right = -0.1] {\tiny 4} (m-1n-1) ;  
        \draw[->] (m1n1.250) -- node[midway, right = 0.1] {\tiny 24} (mn.10) ; 
        \draw[->] (m1n1) -- node[midway, above] {\tiny 23} (mn2) ; 
        \draw[->] (mn2) -- node[midway, above] {\tiny 234} (m-1n1) ;
        \draw[->] (m-1n1) -- node[midway, above] {\tiny 34} (m-2n) ;
    \end{tikzpicture}
    \caption{\label{fig:psu22-short-z1}}
    \end{subfigure}
    \begin{subfigure}{0.48\textwidth}
    \centering
        \begin{tikzpicture}[baseline=(mn.base)]
        \useasboundingbox(-3, -2.2) rectangle (5,2.2);
        \pgfmathsetmacro{\d}{0.7}
        \node[draw=gray, thick, inner sep=2pt] (mn) {$[j_1,j_2]$}; 
        \node[above right = \d of mn] (m1n1) {\scriptsize $[j_1+\frac{1}{2},j_2+\frac{1}{2}]$};
        \node[below right = \d of mn] (m1n-1) {\scriptsize $[j_1+\frac{1}{2},j_2-\frac{1}{2}]$};
        \node[below left = \d of mn] (m-1n-1) {\scriptsize $[j_1-\frac{1}{2},j_2-\frac{1}{2}]$};
        \node[right = {4*\d} of mn] (m2n) {\scriptsize $[j_1+1,j_2]$};
        \node[below = {2*\d} of mn] (mn-2) {\scriptsize $[j_1,j_2-1]$};
        \draw[->] (mn) -- node[midway, below left = -0.1] {\tiny 1} (m1n-1) ; 
        \draw[->] (mn) -- node[midway, above left = -0.1] {\tiny  2} (m1n1) ; 
        \draw[->] (mn) -- node[midway, below right = -0.1] {\tiny 4} (m-1n-1) ; 
        \draw[->] (m1n-1) -- node[midway, above] {\tiny 12} (m2n) ; 
        \draw[->] (m1n-1) -- node[midway, above] {\tiny 14} (mn-2) ;  
        \draw[->] (m2n.250) -- node[midway, below] {\tiny 124} (m1n-1.20) ; 
        \draw[->] (m1n1.250) -- node[midway, right = 0.1] {\tiny 24} (mn.10) ; 
    \end{tikzpicture}
    \caption{\label{fig:psu22-short-z3}}
    \end{subfigure}
    \caption{The decomposition structure of the massive short representations $\expval{j_1,j_2}_{\text{II}\pm}$ with (a) $C = 2\left(j_1 + j_2 + 1\right)$ and (b) $C = -2\left(j_1 + j_2 + 1\right))$.}
    \label{fig:psu22-short-type2}
\end{figure}
For the small values of $j_1 = 0,\frac{1}{2}$ or $j_2 = 0,\frac{1}{2}$, the bosonic components with possibly negative labels would disappear. 
Besides, for $\expval{0,j_2}_{\text{II}+}$, the action of $z_4 z_2$ annihilates the top component.  
The story for the representations satisfying the condition $C = -2\left(j_1 + j_2 + 1\right)$ is very similar. 
The components are annihilated by the action of $z_3$ (Fig.~\ref{fig:psu22-short-z3}), and we call this multiplet $\expval{j_1,j_2}_{\text{II}-}$. 
The special behavior occurs as $j_2=0$, where the top component is also annihilated by $z_4 z_2$ besides the ones generating the negative label components.

\subsection{Infinite-dimensional representations of $\mathfrak{sl}(1,1|2)$}
The supermultiplets appearing in our discussion of the bulk-defect OPE decomposition are all infinite-dimensional given the well-known fact that the unitary representations of the conformal symmetry appearing in the radial quantization of the 2d CFT are all infinite-dimensional. 
In other words, the spacetime $\mathfrak{sl}(1,1)$ part of the bosonic subalgebra $\mathfrak g_0 = \mathfrak{sl}(1,1) \oplus \mathfrak{sl}(2)$ of $\mathfrak{psl}(1,1|2)$ always has infinite-dimensional unitary representations, whereas the R-symmetry $\mathfrak{sl}(2)$ part has finite-dimensional representations. 
The infinite-dimensional $\mathfrak{sl}(1,1)$ representations we consider here are either highest-weight or lowest-weight representations with integer or half-integer conformal weights. 
We call them $\mathcal V(h)$ with $h \neq 0$, which is a lowest-weight representation with the lowest weight $h$ if $h > 0$ or a highest-weight representation with the highest weight $h$ if $h < 0$. 
We label the whole $\mathfrak g_0$ representation as $[\mathcal V(h), j_2]$. 
Recall that the labels $j_1, j_2$ of the finite representations of $\mathfrak{sl}(1,1)$ correspond to their highest weights.
This suggests that the shortening conditions and the classification of the supermultiplets containing the infinite-dimensional highest-weight representation $\mathcal V(-h) \; (h \in \mathbb Z_+/2)$ are simply obtained by replacing the label $j_1 \in \mathbb Z_+/2$ for the finite-dimensional cases with the negative integer $-h$. 
The behaviors of the infinite-dimensional representations are symmetric between $\mathcal V(h)$ and $\mathcal V(-h)$. 
Hence, the classification of the multiplets containing the lowest weight representations $\mathcal V(h)$ is equivalent to that of $\mathcal V(-h)$. 

We comment on some special situations appearing only for the infinite-dimensional multiplets. 
The Type~II conditions can be now satisfied even for the massless $C = 0$ multiplets. 
The top component is either $[\mathcal V(1), 0]$ or $[\mathcal V(-1), 0]$. 
Both of the Type~II conditions are satisfied for this case. 
This means that the top component is annihilated by both $z_1$ and $z_3$, and also $z_4 z_2$ annihilates it since $j_2 = 0$ (Fig.~\ref{fig:psu22-inf-short-type2}). 
\begin{figure}[t]
    \centering
    \begin{subfigure}[b]{0.48\textwidth}
    \centering
        \begin{tikzpicture}
        \pgfmathsetmacro{\d}{0.7}
        \node[draw=gray, thick, inner sep=2pt] (mn) {$[\mathcal V(1), 0]$}; 
        \node[above right = \d of mn] (m1n1) {$[\mathcal V(\frac{1}{2}), \frac{1}{2}]$};
        \draw[->] (mn) -- node[midway, above left = -0.1] {\tiny  2} (m1n1) ; 
    \end{tikzpicture}
    \caption{}
    \end{subfigure}
    \begin{subfigure}[b]{0.48\textwidth}
    \centering
        \begin{tikzpicture}
        \pgfmathsetmacro{\d}{0.7}
        \node[draw=gray, thick, inner sep=2pt] (mn) {$[\mathcal V(-1), 0]$}; 
        \node[above right = \d of mn] (m1n1) {$[\mathcal V(-\frac{1}{2}), \frac{1}{2}]$};
        \draw[->] (mn) -- node[midway, above left = -0.1] {\tiny  2} (m1n1) ; 
    \end{tikzpicture}
    \caption{}
    \end{subfigure}
    \caption{The decomposition structure of the massless infinite-dimensional short multiplets satisfying both of the Type~II shortening conditions.}
    \label{fig:psu22-inf-short-type2}
\end{figure}

\subsection{Bosonic algebra and Casimir invariants \label{sec:bosonic-algebra}}
The conformal symmetry algebra is isomorphic to the direct sum of the two copies of $\mathfrak{sl}(1,1)$ generated by ${(\mathfrak L_i)^\alpha}_{\dot \alpha}$, namely 
\begin{align}
    \{\mathfrak h_1, \mathfrak e_1, \mathfrak f_1\} = \left\{\frac{1}{2}(D+ iM_{34}), P_{1\dot 1}, K_{2\dot 2}\right\}, \quad 
    \{\mathfrak h_2, \mathfrak e_2, \mathfrak f_2\} = \left\{\frac{1}{2}(D- iM_{34}), P_{2\dot 2}, K_{1\dot 1}\right\}
\end{align}
in terms of the bulk $\mathfrak{psl}(2,2|4)$ generators. 
They follow the commutation relations of 
\begin{align}
    [\mathfrak h_i, \mathfrak e_i] = \mathfrak e_i, \quad [\mathfrak h_i, \mathfrak f_i] = - \mathfrak f_i, \quad [\mathfrak e_i, \mathfrak f_i] = -2 \mathfrak h_i. 
\end{align}
Let us think of a lowest-weight representation. 
The lowest-weight state $\ket h$ is the $\mathfrak h_i$ eigenstate with the eigenvalue $h$ and is annihilated by $\mathfrak f = K$. 
$\mathfrak e = P$ raises the weight by one. 
This means that the tensor product of the lowest states in the holomorphic and antiholomorphic parts $\ket{h,\bar h} \equiv \ket h \otimes \ket {\bar h}$ corresponds to the primary state\footnote{To be more precise, we should call it the quasi-primary state since we are arguing the global symmetry of the 2d CFT. } given that it is annihilated by all the special conformal transformation generators. 
The conformal weights of this state are $(h,\bar h)$, giving the dimension and the spin of the state to be $\Delta = h + \bar h = D$ and $s = h-\bar h = i M_{34}$. 
The quadratic Casimir of the representation which $\ket h$ belongs to can be found by acting the Casimir operator on the lowest-weight state. 
On the holomorphic side, the action is found to be 
\begin{align}
    \mathfrak L^2_i \ket h = \left(2 \mathfrak h_i^2 - \mathfrak e_i\mathfrak f_i - \mathfrak f_i \mathfrak e_i\right) \ket h
    = \left(2 \mathfrak h_i^2 - 2\mathfrak h_i\right) \ket h
    = 2h (h-1) \ket h. 
\end{align}
Adding it together with the andiholomorphic side, the Casimir for the whole conformal symmetry is 
\begin{align}\label{eq:casimir-so4}
    (\mathfrak L^2_1 + \mathfrak L^2_2)\ket{h,\bar h}
    = \left(2h (h-1) +2\bar h (\bar h-1)\right)\ket{h,\bar h}
    = (\Delta (\Delta - 2) + s^2 )\ket{h,\bar h}. 
\end{align}
In the Euclidean signature, $P_\mu^\dagger = K_\mu$, in other words $P_{1\dot 1}^\dagger = K_{2\dot 2}$ and $P_{2\dot 2}^\dagger = K_{1\dot 1}$. 
This restricts the value of $h$ to be positive for the unitarity: 
\begin{align}
    0 < |\ket{h + 1}|^2 = \mel{h}{\mathfrak f_i \mathfrak e_i}{h} = \mel{h}{2\mathfrak h_i}{h} = 2h. 
\end{align}
This means that the representations of the conformal symmetry are all infinite-dimensional composed of $\mathcal V(h) \otimes \mathcal V(\bar h)$. 

The story for the R-symmetry $SO(4)_R \simeq SU(2) \times SU(2)$ representation is similar, except that now $\mathfrak e_i^\dagger = -\mathfrak f_i$ for the above commutation relations, and the lowest-weight $m$ is constrained to be negative. 
This means that the unitary representations are finite.






\bibliographystyle{JHEP}
	\cleardoublepage

\renewcommand*{\bibname}{References}

\bibliography{references}
\end{document}